\documentclass{aa}  

\usepackage{graphicx}
\usepackage{txfonts}
\usepackage{cprotect}
\usepackage{orcidlink}
\newcommand{\orcid}[1]{\orcidlink{#1}}

\newcommand{\ec}{{\it Euclid} }
\newcommand{\spitz}{{\it Spitzer} }
\newcommand{\ie}{I$_{\rm E}$}
\newcommand{\ye}{Y$_{\rm E}$}
\newcommand{\je}{J$_{\rm E}$}
\newcommand{\he}{H$_{\rm E}$}
\newcommand{\gaia}{{\it Gaia} }

\begin{document}

   \title{Free-floating planetary mass objects in LDN~1495 from Euclid Early Release Observations}

   \author{H. Bouy\orcid{0000-0002-7084-487X}\inst{1, 2}
          \and
          E. L. Mart\'\i n\orcid{0000-0002-1208-4833}\inst{3} 
          \and
           J.-C. Cuillandre\orcid{0000-0002-3263-8645} \inst{4}
          \and D. Barrado\orcid{0000-0002-5971-9242}\inst{5}
          \and M. Tamura\orcid{0000-0002-6510-0681} \inst{6,7}
          \and E. Bertin\orcid{0000-0002-3602-3664} \inst{4}
          \and M.~{\v Z}erjal\orcid{0000-0001-6023-4974}\inst{3,8}
          \and S. Points\orcid{0000-0002-4596-1337}\inst{9}
          \and J. Olivares\orcid{0000-0003-0316-2956}\inst{10}          
          \and N. Huélamo\orcid{0000-0002-2711-8143}\inst{5}
          \and T. Rodrigues\orcid{0009-0002-0946-6969}\inst{1}
          }

   \institute{Laboratoire d'astrophysique de Bordeaux, Univ. Bordeaux, CNRS, B18N, allée Geoffroy Saint-Hilaire, 33615 Pessac, France.\\
              \email{herve.bouy@u-bordeaux.fr}
         \and
         Institut universitaire de France (IUF), 1 rue Descartes, 75231 Paris CEDEX 05
         \and          
         Instituto de Astrof{\'{\i}}sica de Canarias, E-38205 La Laguna, Tenerife, Spain 
         \and
         Universit\'e Paris-Saclay, Universit\'e Paris Cit\'e, CEA, CNRS, AIM, 91191, Gif-sur-Yvette, France
         \and
         Centro de Astrobiología (CAB), CSIC-INTA, ESAC Campus, Camino bajo del Castillo s/n, E-28692 Villanueva de la Ca\~nada, Madrid, Spain 
         \and
         Department of Astronomy, Graduate School of Science, The University of Tokyo, Tokyo, Japan
         \and 
         National Astronomical Observatory of Japan,  Tokyo, Japan
         \and 
         Universidad de La Laguna, Dpto. Astrof{\'{\i}}sica, E-38206 La Laguna, Tenerife, Spain
        \and
         Cerro Tololo Inter-American Observatory, NSF’s NOIRLab, Casilla 603, La Serena, Chile
         \and
         Departamento de Inteligencia Artificial, Universidad Nacional de Educación a Distancia (UNED), c/Juan del Rosal 16, E-28040, Madrid, Spain
             }

   \date{Received ; accepted 22 Feb 2025}

 
  \abstract
   {Substellar objects, including brown dwarfs and free-floating planetary-mass objects, are a significant product of star formation. Their sensitivity to initial conditions and early dynamical evolution makes them especially valuable for studying planetary and stellar formation processes.}
   {We search for brown dwarfs and isolated planetary mass objects in a young star-forming region to better constrain their formation mechanisms. }
   {We took advantage of the \ec unprecedented sensitivity, spatial resolution and wide field of view to search for brown dwarfs and free-floating planetary mass objects in the \object{LDN 1495} region of the Taurus molecular clouds. We combined the recent \ec Early Release Observations with older very deep ground-based images obtained over more than 20~yr to derive proper motions and multiwavelength photometry and to select members based on their morphology and their position in a proper motion diagram and in nine color-magnitude diagrams.}
   {We identified 15 point sources whose proper motions, colors, and luminosity are consistent with being members of LDN~1495. Six of these objects were already known M9–L1 members. The remaining nine are newly identified sources whose spectral types might range from late-M to early-T types, with masses potentially as low as 1$\sim$2 M$_{\rm Jup}$ based on their luminosity and according to evolutionary models. However, follow-up observations are needed to confirm their nature, spectral type, and membership. When it is extrapolated to the entire Taurus star-forming region, this result suggests the potential presence of several dozen free-floating planetary mass objects.
}
   {}


   \maketitle
%

\section{Introduction}
The Taurus molecular clouds (TMC) have been a focal point for studies of star formation. They provide invaluable insights into the processes that lead to the birth and early evolution of stars and substellar objects \citep[see ][ and references therein]{2008hsf1.book.....R}. Their youth \citep[1 to 10~Myr,][]{Luhman2023b} and proximity \citep[between 120 and 200~pc,][]{Galli2019} make them an ideal target for observational studies, and they allow detailed observations of all stages of the formation of stars, from dense molecular cores \citep{Marsh2016}, filamentary structures \citep{Hacar2013,Palmeirim2013,Roccatagliata2020,Kirk2024}, and embedded protostars that are surrounded by dense envelopes \citep{Andre1999} to more evolved Class II young stellar objects (YSOs)\footnote{including TTauri stars, which are named after the prototype, T Tauri, which is located in the TMC}  with circumstellar disks and their forming planetary systems \citep[e.g. ][ and references therein]{Rebull2010,Ricci2010,Yamaguchi2024,Garufi2024}, to the final products of star formation. 

The TMC exhibit certain characteristics that distinguish them from other star-forming regions. The scarcity of massive stars, and consequently, the minimum feedback influencing the dynamics and turbulence of molecular clouds, makes Taurus an exceptional laboratory for investigating the fundamental processes of star formation and the role of feedback (or lack thereof) from massive stars.

Substellar objects, including brown dwarfs and free-floating planetary-mass objects, are expected to be particularly insightful in this context, as their formation is highly sensitive to the conditions of their parent molecular cloud (turbulence and feedback) and the early dynamical evolution within the proto-cluster, both of which vary very much when massive stars are present. This  has made Taurus a popular place for searches for brown dwarfs and free-floating planetary-mass objects  \citep{Itoh1996, Tamura1998, Martin2001, Briceno2002,Luhman2006a, Luhman2004, Luhman2023b, Guieu2006, Esplin2017, Esplin2019}. Several scenarios have been proposed to explain their formation. In the first scenario, they form like stars, through the contraction and collapse of a small molecular core. In other scenarios, they are ejected or stripped from their parent formation site before they can accrete enough material to become very low-mass stars. Each scenario results in different predictions regarding the properties of these objects, particularly in terms of their expected numbers and the possible existence of a minimum mass below which these low-mass objects cannot form.

Moreover, sensitive mid- and far-infrared surveys have enabled detailed censuses of protoplanetary disks around  Taurus substellar objects \citep{Luhman2006b, Rebull2011, Esplin2014, Esplin2019, Scholz2006} and facilitated the identification of  candidates of class I proto-brown dwarfs \citep{Barrado2009, Palau2012, Morata2015}. These advancements highlight Taurus as a prime region for exploring the formation and early evolution of substellar objects.

We combine \ec images with deep multi-epoch ground-based images of the LDN 1495 cloud, which is located at 130~pc \citep{Galli2019},  to search for ultracool  dwarf (UCD) members and set new constraints on the formation of these elusive objects.  The \ec observations, combining unprecedented depth with a wide field of view and a high spatial resolution, extend previous studies that were based on ground-based data and {\it Gaia} DR3. They also serve as a valuable complement to Jame Webb Space Telescope (JWST) surveys, which, while providing exceptional sensitivity, are limited to much smaller regions.

\section{Euclid observations and data reduction}
\begin{figure}
   \centering
   \includegraphics[width=0.49\textwidth]{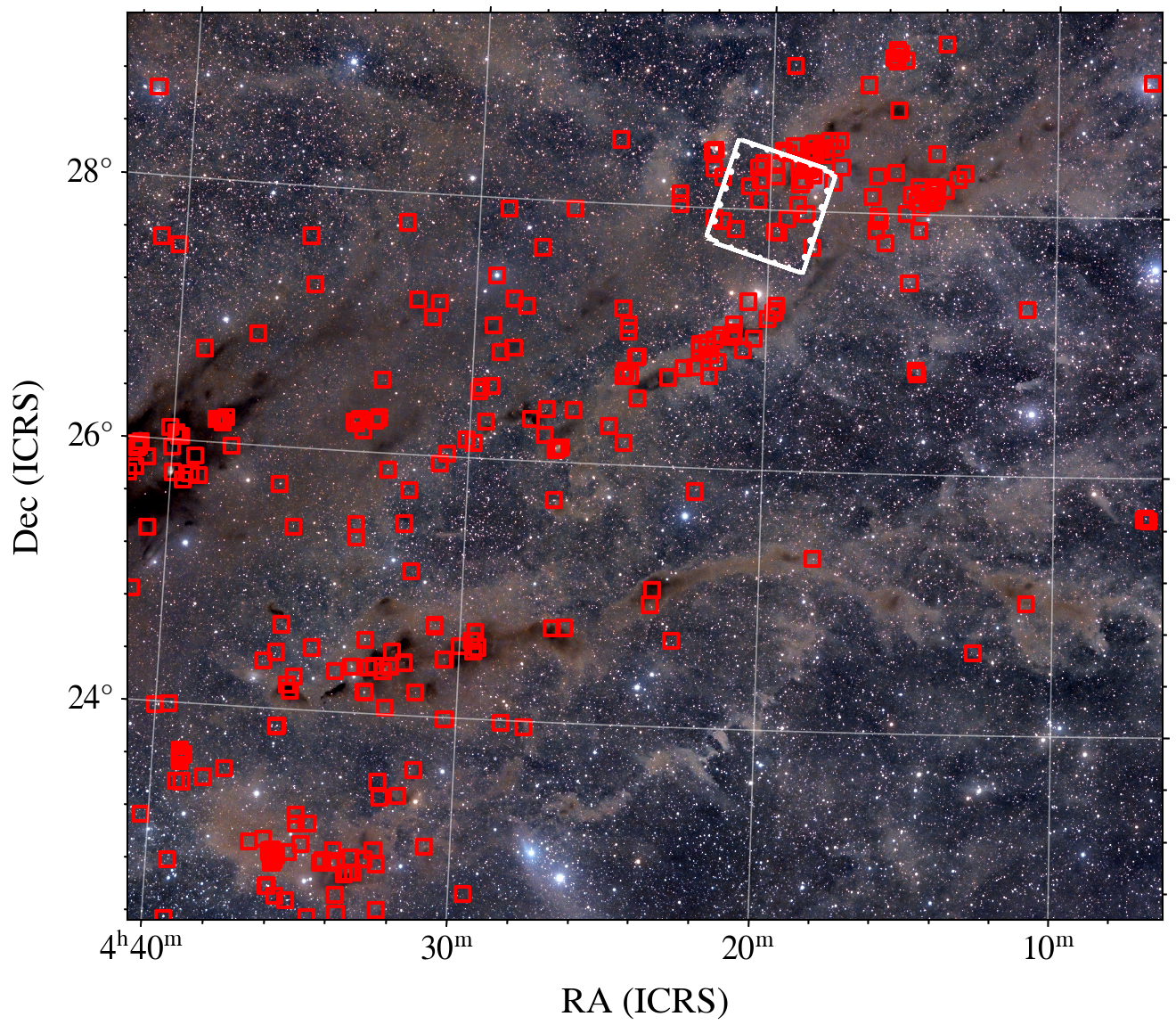}
   \caption{Photograph of the LDN1495 region showing the Taurus molecular clouds and the Taurus members presented by \citet{Galli2019} and \citet{Esplin2019} as red squares. The \ec observation field of view is represented by a white square. Photograph credit: Bobby White}
              \label{fig:pointing}%
    \end{figure}

A field of the TMC centered near the \object{LDN 1495} cloud on $\alpha_{\rm J2000}=$04h19m56s and $\delta_{\rm J2000}=$+28\degr01\arcmin23.9\arcsec and covering approximately 1\degr$\times$1\degr\, was observed with the European Space Agency (ESA) \ec observatory as part of the Early Research Observation (ERO) program on 2023 September 16 following the standard \ec observational strategy described by \citet{2024arXiv240513491E}. The program and observations are described in detail by \citet{Martin2024}, and Fig.~\ref{fig:pointing} shows the footprint of the observations in the TMC. 

The data processing was described by \citet{Cuillandre2024}, and we used the stacked mosaics for the present study. The background in the stacked images was modeled and subtracted using DeNeb, a new software designed to remove nebulae and extended emission from astronomical images using deep-learning (Bertin et al., in prep.). 

\ec has two scientific instruments, the visible band imager VIS \citep{VIS}  covering 0.787\degr$\times$0.709\degr\, with a single broad band from 550~nm to 950~nm and a pixel scale of 0\farcs1,  and the Near-Infrared Spectrometer and Photometer \citep[NISP,][]{NISP} with three filters \citep[\ye, \je, \he, see Table~3 of ][]{NISP_FILTERS} that cover roughly the same area as VIS with a pixel scale of 0\farcs3. An overview of the filters main properties is given in Table~\ref{table:euclidfilters}. For the NISP images, all the sources were extracted with the \verb|SExtractor| package \citep{Bertin1996} using a $\chi^2$ image made of a linear combination of the \ye, \je, and \he\, images following the method described by \citet{1999AJ....117...68S} and optimized for simultaneous multiband detection of faint objects. For VIS, we extracted all the sources with more than three pixels above 1.5 standard deviation of the local background. In all cases, we then measured the positions using the point spread function (PSF) and Sérsic model-fitting option in \verb|SExtractor| , which relies on the empirical PSF model previously derived by \verb|PSFex| \citep{Bertin2011}. The Sérsic model-fitting  offers several advantages for our purpose. The PSF-convolved Sérsic model fit offers a level of astrometric accuracy that is comparable with the level of pure PSF fits for point sources while making it possible to measure precise galaxy positions and photometry. It also delivers morphometric parameters that allow us to identify extended objects such as galaxies, using in particular the \verb|SPREAD_MODEL| parameter \citep[see][ and discussion below]{Bouy2013}. This proves particularly useful in extremely deep and sharp images that are largely dominated by extragalactic objects. Finally, model-fitting parameters are largely immune to the spatial discretization effects caused by undersampling, which is crucial in the case of NISP and its undersampled PSF. We used the Kron aperture photometry reported by \verb|SExtractor|  as \verb|MAG_AUTO| and converted from instrumental into apparent magnitudes using the nominal  AB flux zeropoints provided by the ERO pipeline. The \ec VIS images saturate around \ie$\sim$18~mag, the NISP \ye\, and \je\, images around 16~mag, and the NISP \he\, image around 16.5~mag, and we discarded detections below these values. A total of 27 members from the combined lists of \citet{Esplin2019} and \citet{Galli2019} fall within the \textit{Euclid} images. All are detected at least in one \ec band; however, 10 of them are saturated in one or several images and were therefore discarded from our analysis. 

\section{Complementary observations}
We complemented the \ec images and catalogs with images obtained at various space and ground-based facilities gathered in the context of the COSMIC-DANCE survey of the TMC \citep[][]{Bouy2013}. They include in particular (see also Table~\ref{tab:observations})
\begin{itemize}
\item CFHT/CFH12K images in the \textit{I} band
\item CFHT/MegaCam images in the \textit{r}, \textit{i}, and \textit{z} bands
\item Subaru SuprimeCam images in the \textit{i} and \textit{z} bands
\item Subaru HSC images in the r, \textit{i},  \textit{z}, and \textit{y}-bands
\item CFHT/WIRCam images in the \textit{J},  \textit{H}  , and \textit{W} bands
\item UKIRT/WFCAM images in the \textit{Z}, \textit{Y},  \textit{J},  \textit{H}  , and \textit{Ks} bands
\item KPNO/NEWFIRM images in the \textit{J},  \textit{H}  , and \textit{Ks} bands
\item CAHA3.5/O2000 images in the \textit{H}   and \textit{Ks} bands
\item JAST/T80cm images in the \textit{i} and \textit{z} bands.
\end{itemize}
More details about these instruments and observations are given in Table~\ref{tab:observations}.

These images were processed using the procedure described by \citet{Bouy2013} and \citet{Miret2022}, and we applied DeNeb to subtract any extended emission, and the detected sources and their PSF and Sérsic model-fitting parameters were measured using the same procedure as above. The photometric calibration of the \textit{i},  \textit{z}, and \textit{y} band images was tied to Pan-STARRS \citep{ps1}, the calibration of the \textit{J},  \textit{H} ,  and \textit{Ks} images to 2MASS \citep{2MASS}, and that of the \textit{Z}  and \textit{Y} images to the UKIDSS survey \citep{UKIDSS}. The photometric calibration of the CFHT/UH12K \textit{I} band and WIRCAM \textit{W} band was not attempted.

We also retrieved all the Level 1 BCD images acquired with the {\it Spitzer Space Telescope} \citep{Spitzer} and its IRAC camera \citep{IRAC}, which overlap with the \ec field of view from the {\it Spitzer Heritage Archive}. The images come from the \spitz Taurus survey, which was conducted by \citet[programs 30816 and 3584]{Padgett2007}, and from the {\it Spitzer} Heritage Project by \citet[program 462]{Rebull2010}. We processed them following the recommended procedure using the latest version of MOPEX \citep[v18.5.0;][]{MOPEX}, and we applied DeNeb on the final mosaics to remove the extended emission from the cloud. The sources were then detected and their positions and fluxes measured using \verb|SExtractor| and the same strategy as for the \ec and ground-based images mentioned above, and the uncertainty maps delivered by the MOPEX pipeline were used as weight maps. 

Our Spitzer catalog sensitivity is significantly better than that of \citet{Rebull2010} because we combined all available epochs, applied DeNeb to remove the bright nebula, and had less restrictive source-detection parameters. The photometry was tied to their photometry, and we estimate that the catalog is complete up to \textit{I1}/\textit{I2}$\sim$18.0~mag and sensitive to \textit{I1}/\textit{I2}$\sim$19.5~mag. This provides a very good overlap with Euclid down to the ultracool dwarf regime.

\section{Proper motions}

\begin{figure}
   \centering
   \includegraphics[width=0.49\textwidth]{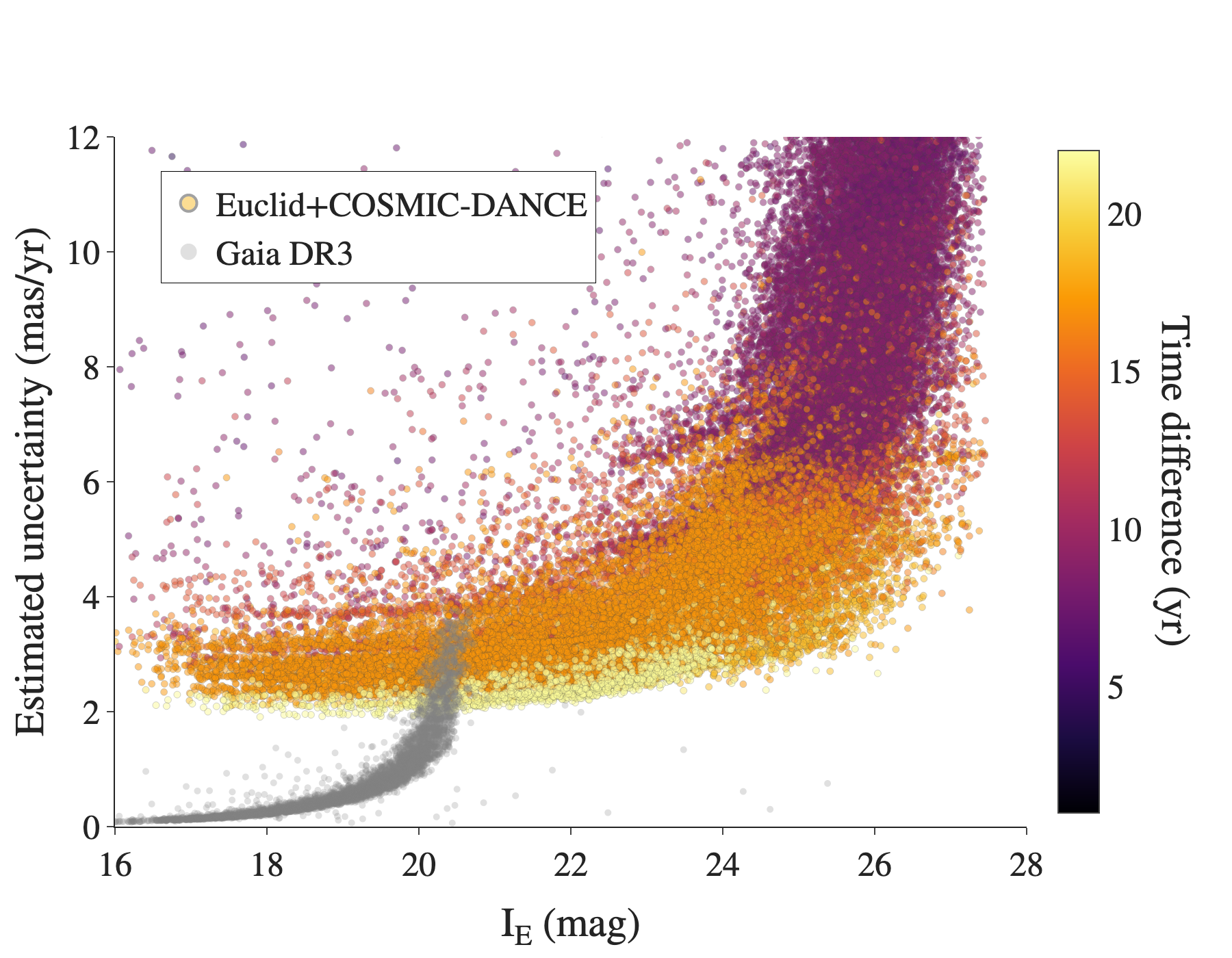}
   \caption{Estimated uncertainties on the proper motions vs \ie\, magnitude for all the sources in the Euclid field. The color scale represents the time difference between the \ec and the earliest observations available for a source. The \gaia\, DR3 catalog over the same area is overplotted as gray dots.}
   \label{fig:ppm_uncertainties}%
    \end{figure}

Proper motions were derived following the method described by \citet{Bouy2013} and used all the above-mentioned data except for the \spitz data, which have a much coarser spatial resolution. Figure~\ref{fig:ppm_uncertainties} shows the estimated uncertainty as a function of the apparent \ie\  magnitude, as well as a comparison with \gaia\, DR3 measurements for sources detected in the \ec images. Uncertainties as low as 3$\sim$4~mas yr$^{-1}$ were achieved up to \ie$\sim$26~mag when the time baseline is long, thanks in particular to the very deep 2015 and 2016 Subaru HSC images, which match the VIS depth and provide a time baseline of 7-8 years.

As explained by \citet{Bouy2013}, the proper motions we computed are relative to one another and display an offset with respect to the geocentric celestial reference system. We estimated the offsets by computing the median of the difference between our proper motions and those reported in \gaia\, DR3 and applied the corresponding correction to all our measurements. The \ec VIS images saturate around \ie$\sim$18~mag, the \ye\, and \je\, around 16~mag, and the \he\, around 16.5~mag. This leaves an overlap of about 3~mag with the \gaia\, DR3 catalog \citep{Gaia2016,Gaia2023} and ensures a robust and precise estimate of the correction.  

\section{Member selection}
The number count of extragalactic sources in an astronomical survey increases geometrically with the sensitivity of  the observations. It reaches almost one million of extragalactic sources per square degree at the VIS sensitivity limit   \citep[see e.g.][]{Capak2007}. The ultracool dwarf density in Taurus over the same magnitude range still remains to be measured, but is expected to be many orders of magnitudes lower. This makes their identification very difficult among an  overwhelming number of extragalactic sources that can have similar luminosities and colors in the \ec bandpasses.

The strategy we chose to select ultracool dwarf candidates involved a selection using multiple criteria, which included
\begin{enumerate}
\item the morphology of the sources to reject extended sources that must be (mostly) extragalactic
\item the proper motion of the sources. The extragalactic sources motions should not be detectable with the precision achieved by our measurements, while the motion of LDN~1495 members is well known and large enough \citep[$\sim$30 mas yr$^{-1}$,][]{Galli2019, Luhman2023b} to be clearly detected in most cases
\item the location of the sources in nine color-magnitude diagrams, including optical and near-infrared photometry
\item a visual inspection of the images to discard problematic sources
\end{enumerate}

This strategy offered the advantage of identifying robust members, but the inconvenience of missing some genuine members (e.g., barely resolved binaries; objects with missing or poor proper motion measurements; and objects missing some photometric measurements, e.g., deeply embedded sources or objects with extended emission related to e.g. disks or outflows). Our main objective was therefore to identify reliable ultracool candidates and not to perform a complete and unbiased census of ultracool dwarfs in the \ec field of view. This will be done in a future analysis.

\subsection{Selection on the morphology}

\begin{figure}
   \centering
   \includegraphics[width=0.49\textwidth]{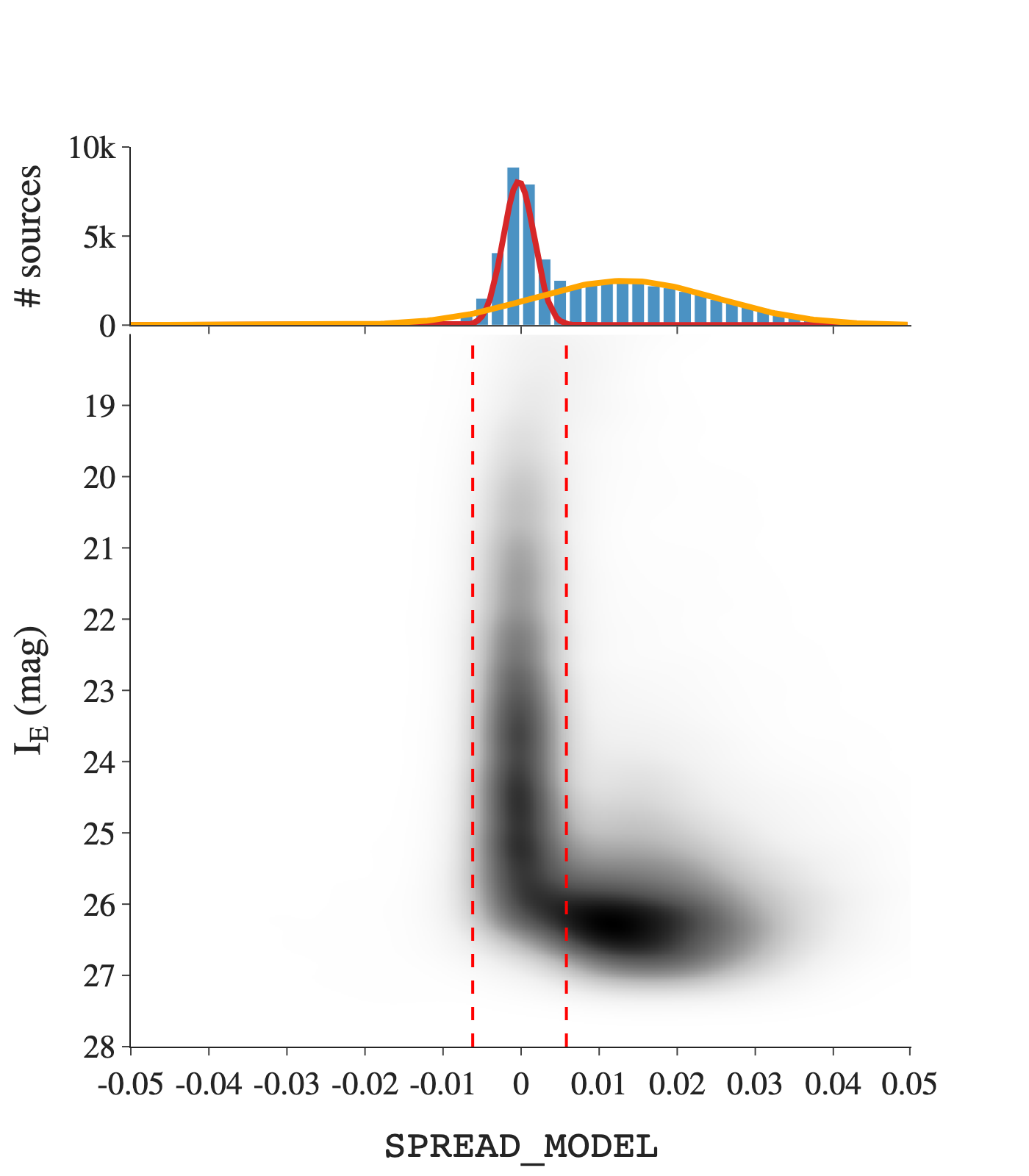}
   \cprotect\caption{Distribution (kernel density estimate) of the \verb|SPREAD_MODEL| as a function of \ie\,  magnitude. Two normal distributions were adjusted to the histogram of \verb|SPREAD_MODEL| (upper panel). The Gaussian fits to the histogram of the distribution are represented in red and orange. The dashed vertical lines indicate the 3$\sigma$ limits we chose to distinguish between point sources and extended sources. }
   \label{fig:morphometry}%
    \end{figure}

Figure~\ref{fig:morphometry} shows the distribution of  \verb|SPREAD_MODEL| in the VIS images as a function of VIS \ie\, magnitude. As described by \citet{Bouy2013}, point sources are located in a sequence centered around zero in this diagram, and  sources that deviate significantly from a point source are located to the left (mostly cosmic-ray hits, which are rare in these mosaics) or to the right (mostly extended extragalactic sources or diffuse emission, the latter are greatly reduced by DeNeb). 

As illustrated in Fig.~\ref{fig:morphometry}, the \verb|SPREAD_MODEL| distribution can be nicely modeled by the sum of two Gaussian distributions, one for the point-sources and one for the extended sources at higher positive \verb|SPREAD_MODEL| values. We defined the locus of point sources between -0.006$\le$\verb|SPREAD_MODEL|$\le$0.006, which corresponds to $\pm3\sigma$ of the associated Gaussian fit, and discarded all the sources outside this domain. 

This selection unfortunately discards genuine blended visual binaries that are unresolved by \verb|SExtractor| (with a separation of about the diffraction limit, i.e., around 0\farcs2 for VIS) or genuine young stellar objects with extended emission from a disk, envelope, outflows, or jets, but it is expected to remove a significant fraction of extragalactic sources. According to cosmological models, the angular size of galaxies indeed decreases to $\lesssim$1\arcsec\, around $z\sim1$ and then increases beyond \citep[e.g.][]{Hoyle1959, Peebles1993}, and a significant fraction of galaxies should therefore be easily resolved in \ec VIS and NISP images.

\subsection{Selection in proper motion \label{sec:ppm}}
\begin{figure}
   \centering
   \includegraphics[width=0.49\textwidth]{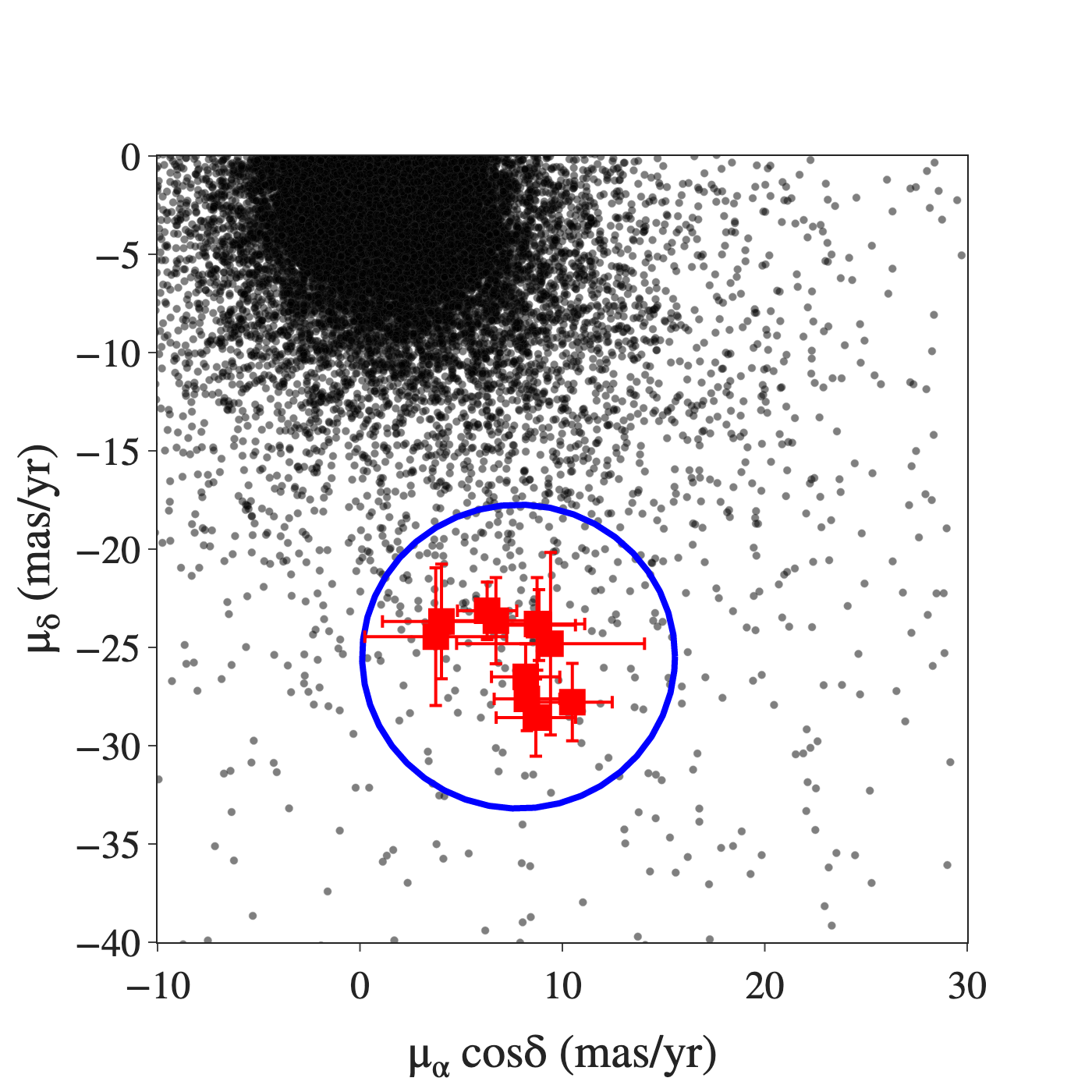}
   \caption{Proper motion diagram for the sources in our catalog, selected based on their morphology. The members from \citet{Galli2019} and \citet{Esplin2019} with a counterpart in our catalog are represented as red squares. A blue circle shows the area we used for the member selection.}
   \label{fig:ppm_selection1}%
    \end{figure}

Figure~\ref{fig:ppm_selection1} shows the proper motion diagram for the sources that remained after the \verb|SPREAD_MODEL| criterion described above was applied.  Of the 27 known Taurus members  reported in either \citet{Galli2019} or \citet{Esplin2019}, only 17 have a counterpart in our catalog.  Of these 17 members, 11 have a reliable proper motion measurement in our catalog. The other 6 have either very large uncertainties ($\gtrsim$10~mas yr$^{-1}$) or clearly incorrect values compared to Gaia DR3. This is related to issues in their individual proper motion determination. These 11 known members are represented in Fig.~\ref{fig:ppm_selection1}, and we computed their weighted mean motion and associated weighted standard deviation as ($\mu_{\alpha}\cos{\delta},\mu_{\delta})=(7.8,-25.5)\pm(2.5,2.5)$~mas yr$^{-1}$. We then selected sources with estimated uncertainties on the proper motion that were smaller than 10~mas yr$^{-1}$ and that were located within 3$\sigma$ of this mean motion as member candidates, as illustrated in Fig.~\ref{fig:ppm_selection1}. Out of these 176 sources, 5 are detected in fewer than four filters, and we discarded them because such a small number of photometric measurements is not enough to infer their membership. This left a sample of 171 objects.

\subsection{Selection in color-magnitude diagrams}
\begin{figure*}
   \centering
   \includegraphics[height=0.95\textwidth]{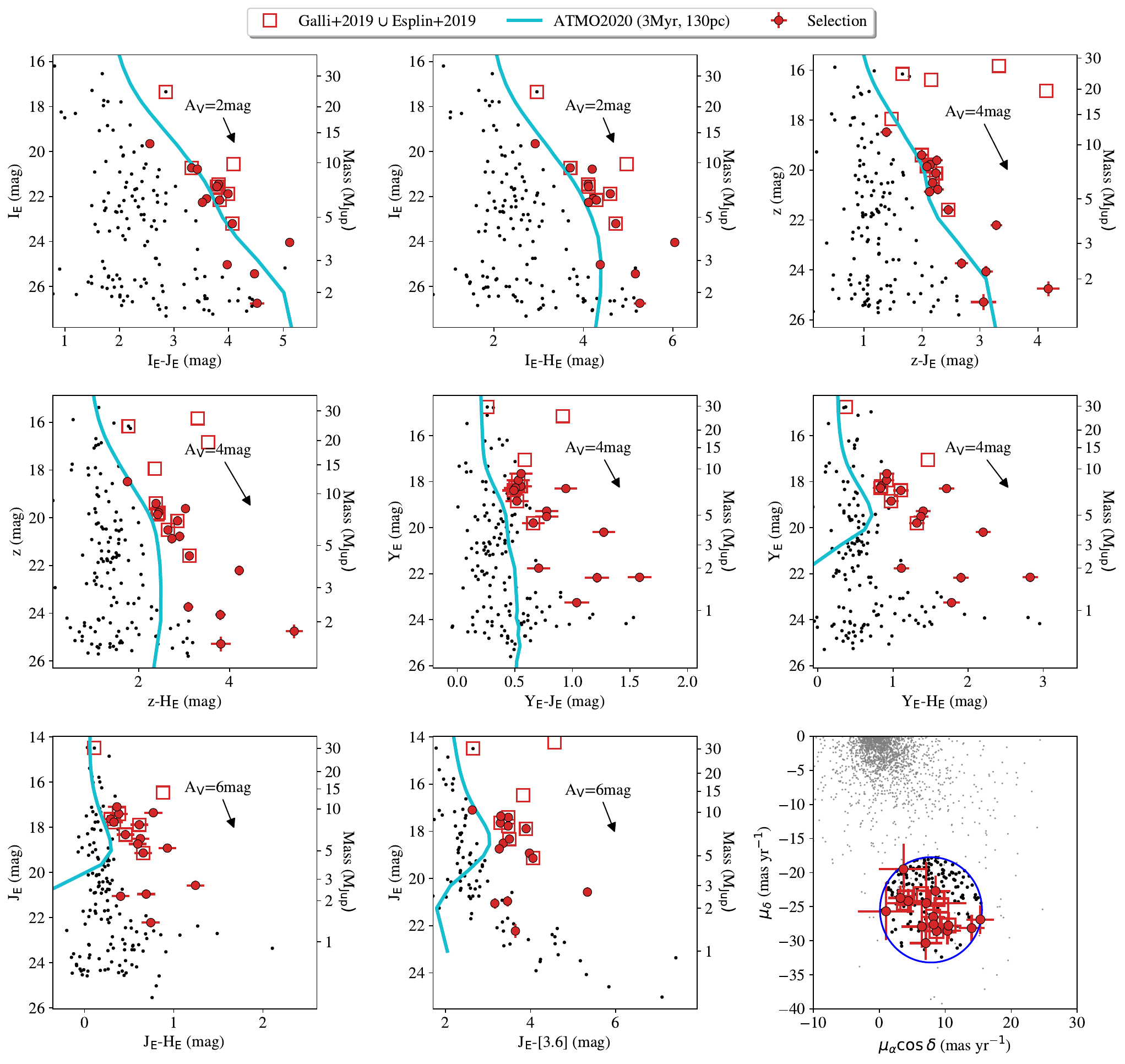}
   \cprotect\caption{Color-magnitude and proper motion diagrams used for the candidate selection. The sources selected using the \verb|SPREAD_MODEL| and proper motion criteria described in the text are represented with black dots. The 15 objects selected in this work are represented with red dots. Known Taurus members from \citet{Galli2019} and \citet{Esplin2019} present in the \ec images are overplotted with red open squares. The \citet{Chabrier2023} isochrone for 3~Myr and 130~pc is represented by a light blue curve. The corresponding mass scale is indicated on the right axis. Extinction vectors are also represented. }
   \label{fig:cmd-models}%
    \end{figure*}

 To refine the previous selection, we analyzed the positions of the 171 selected sources in various color-magnitude diagrams. Ideally, an empirical sequence of confirmed ultracool members would be used to define selection limits in these diagrams. However, no such sample exists in the \ec bandpasses, as known Taurus members falling in the \ec images reach only \ie$\sim$23~mag. Although theoretical isochrones are known to show some discrepancies with observations, in particular at these young ages, we opted to compare the luminosities and colors of previous candidates with predictions from the \citet{Chabrier2023} models for 3 Myr at a distance of 130 pc  \citep{Galli2019}. The selection process was sequential, starting with (\ie, \ie-\je), followed by (\ie, \ie-\he), (\ye, \ye-\je), (\ye, \ye-\he), and finally, (\je, \je-\he), as shown in Fig.~\ref{fig:cmd-models}. At each stage, candidates were selected when their positions were on or to the right of the isochrone within their 3$\sigma$ uncertainties. To account for dispersion in parameters such as distance or youth-related excesses, the isochrone was shifted 0.1 mag to the left. After this process, nine candidates were identified. 

This initial selection revealed that the \ie, band limits the candidate detection at the faint end. To address this, we performed a second, independent selection by replacing the \ie, band with our ground-based {\it z}-band photometry. A similar sequential process was then applied to identify candidates within this adjusted parameter space, starting with (\textit{z}, \textit{z}-\je), followed by (\textit{z}, \textit{z}-\he), (\ye, \ye-\he), and finally, (\je, \je-\he). This approach identified 15 candidates, 8 of which overlapped with the initial selection. The remaining missing candidate was assessed and was excluded due to its anomalously blue colors in (\textit{z}-\je) and (\textit{z}-\he), and its luminosity in the \textit{r}-band was also incompatible. This resulted in a final sample of 15 candidates from both selections. Six of these candidates were already reported in \citet{Esplin2019} and classified as M9-L1 members, which added confidence to our selection and to the nature of the 9 new candidates. Interestingly, the candidates that were selected using the {\it z}-band all fall slightly blueward the theoretical isochrone in the (\ie, \ie-\je) diagram. This suggests that the models might miss important features in these bands. 

\begin{figure*}
   \centering
   \includegraphics[height=0.95\textwidth]{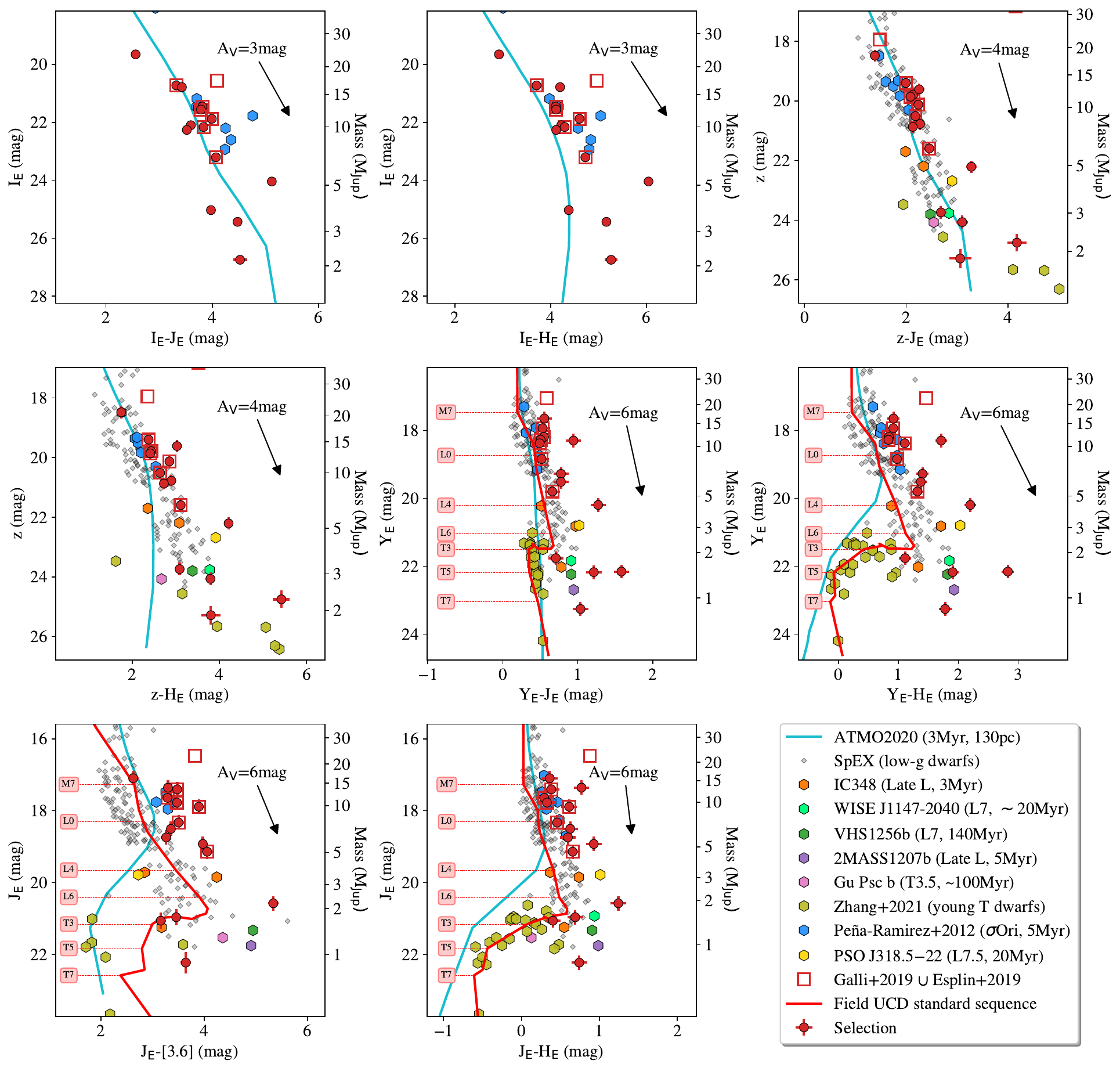}
   \caption{Color-magnitude diagrams for the 15 selected objects (same symbols as in Fig.~\ref{fig:cmd-models}) and known young ultracool dwarfs in IC348 from \citet[orange]{Luhman2024}, young T-dwarfs from \citet[olive green]{Zhang2021}, the young L7 companion VHS 1256b \citep[green;][]{Miles2023}, the young late-L companion 2MASS~1207b \citep[purple;][]{Luhman2023a}, the young T3.5 dwarf companion Gu Psc b \citep[pink;][]{Naud2014}, the young L7 dwarf WISEA~J114724.10-204021.3 \citep[cyan;][]{Schneider2016}, the young L7 dwarf PSO J318.5338-22.8603 \citep[yellow;][]{Liu2013}, and all the ultracool dwarfs classified as low or intermediate gravity in the IRTF/SpeX ultracool dwarf library \citep[grey diamonds;][]{SPLAT}. Their photometry in the Euclid bands was derived from their spectra as described in the text. The SPLAT standard sequence for field ultracool dwarfs is also represented as a red line when the corresponding filters are available, and the spectral type scale is indicated. Extinction vectors are also represented.}
   \label{fig:cmd-knownucd}%
    \end{figure*}

To further validate our selection based on uncertain theoretical isochrones, we plotted the 15 candidates in various color-magnitude diagrams to compare their luminosities and colors with those of known ultracool dwarfs (Fig.~\ref{fig:cmd-knownucd}). One of the initial challenges encountered in this analysis was the lack of benchmark photometric measurements for ultracool objects in the \ec  filters. The \ec  VIS and NISP filters differ significantly from other standard instruments and systems. Therefore, we undertook the task of building a library of spectro-photometric measurements to guide our search for ultracool objects in the \ec  photometric system, as described in Appendix~\ref{sec:spectrophoto} and using the \ec filter properties described in Appendix~\ref{sec-euclidfilters}. We also included \ec  photometry extracted from the Early Release Observations presented by \citet{Martin2024} for candidate  members of the $\sigma$-Orionis cluster from  \citet{Pena2012}.

Figure~\ref{fig:cmd-knownucd} shows that, as expected, known young ultracool dwarfs display a wide range of colors and luminosities. Additionally, it shows a well-known discrepancy around the late-L to early-T transition, where models turn prematurely blueward in near-infrared color-magnitude diagrams instead of continuing redward, as observed in the young ultracool population. This deviation is attributed to the insufficient incorporation of dust opacity in these models \citep[][ see also Fig.~\ref{fig:cmd_known_ucd}]{Marley2021}. However, this discrepancy is expected to have only a moderate impact on our selection, as we specifically targeted sources positioned redward of the isochrones.

Our nine new candidates exhibit properties that are consistent with late-M to early-T dwarfs, as illustrated in Fig.~\ref{fig:cmd-models} and \ref{fig:cmd-knownucd}. This places them well below the planetary mass limit for the age \citep[1$\sim$2~Myr,][]{Luhman2023b} and distance \citep[130~pc][]{Galli2019} of LDN~1495. It is also possible that some of these candidates are objects with earlier spectral types embedded within the cloud, as extinction moves object redward in all these diagrams. Follow-up spectroscopy is required to confirm their nature and determine precise spectral types.

If these nine objects are confirmed as FFPs and this result is extrapolated to the entire Taurus molecular cloud region ($\sim$100 deg$^{2}$), it suggests the potential presence of several dozen free-floating  planetary mass objects in this region.

\subsection{Visual inspection and spatial distribution}

\begin{figure}
   \centering
   \includegraphics[width=0.45\textwidth]{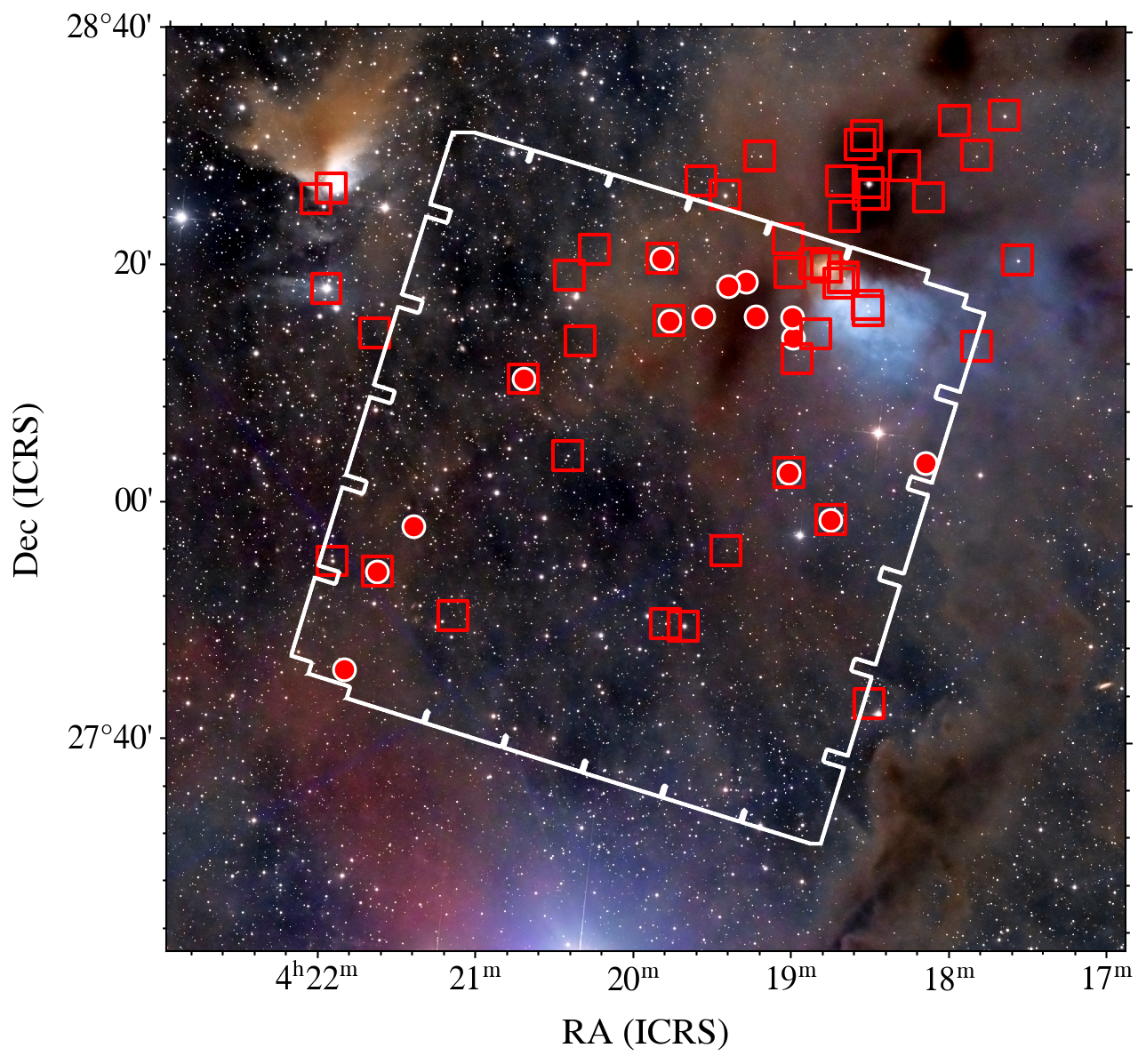}
   \caption{Spatial distribution of our candidates (red circles) and of known members from \cite{Esplin2019} and \citet[red crosses]{Galli2019}. The \ec  VIS image field ofview is represented as a white rectangle. Photograph credit: Chris Fellows.}
   \label{fig:positions}%
    \end{figure}

We visually inspected the VIS and NISP images for each of the 15 candidates and verified that all detections were consistent and reliable. The astrometric, photometric, and morphometric measurements are given in Table~\ref{tab-candidates}. Finally, Fig.~\ref{fig:positions} shows the spatial distribution of the 15 objects we selected  and compares them to the distribution of known members from \citet{Galli2019} and \citet{Esplin2019}. Interestingly, these samples appear to be similarly distributed, with most candidate members located in the northwestern part of the field.

\begin{table}
\caption{Candidate astrometry and photometry.}
\label{tab-candidates}
\begin{tabular}{lcc}
\hline\hline
Object & RA (J2000) & Dec (J2000)  \\
DANCe  & (deg) & (deg)  \\
  \hline
J041808.97+280336.5 & 64.53739 & 28.06013 \\ 
J041845.30+275847.8\tablefootmark{a} & 64.68875 & 27.97994 \\
J041859.75+281410.4 & 64.74897 & 28.23622 \\
J041900.12+281551.7 & 64.75052 & 28.26437 \\
J041901.28+280248.2\tablefootmark{a} & 64.75534 & 28.04671 \\
J041914.29+281556.6 & 64.80956 & 28.26574 \\
J041917.96+281851.5 & 64.82484 & 28.3143 \\
J041924.83+281829.3 & 64.85345 & 28.30815 \\
J041934.21+281558.3 & 64.89253 & 28.26619 \\
J041947.39+281534.5\tablefootmark{a} & 64.94745 & 28.25959 \\
J041950.43+282048.4\tablefootmark{a} & 64.96013 & 28.34679 \\
J042043.02+281036.1\tablefootmark{a} & 65.17927 & 28.1767 \\
J042124.70+275805.9 & 65.3529 & 27.96831 \\
J042138.50+275414.4\tablefootmark{a} & 65.41042 & 27.904 \\
J042150.63+274556.3 & 65.46095 & 27.76564 \\
  \hline
\end{tabular}
\tablefoottext{a}{Previously known member}
\tablefoot{The complete table including proper motions and photometric measurements is available in electronic form}
\end{table}

\section{Previously known members or candidates}
A number of previously known members or candidate members of the Taurus association fall within the \ec images. Most are saturated and do not appear in our catalog, but we discuss a few famous objects and proto-brown dwarf candidates.

\subsection{Landmark objects}
\begin{figure}
   \centering
   \includegraphics[width=0.45\textwidth]{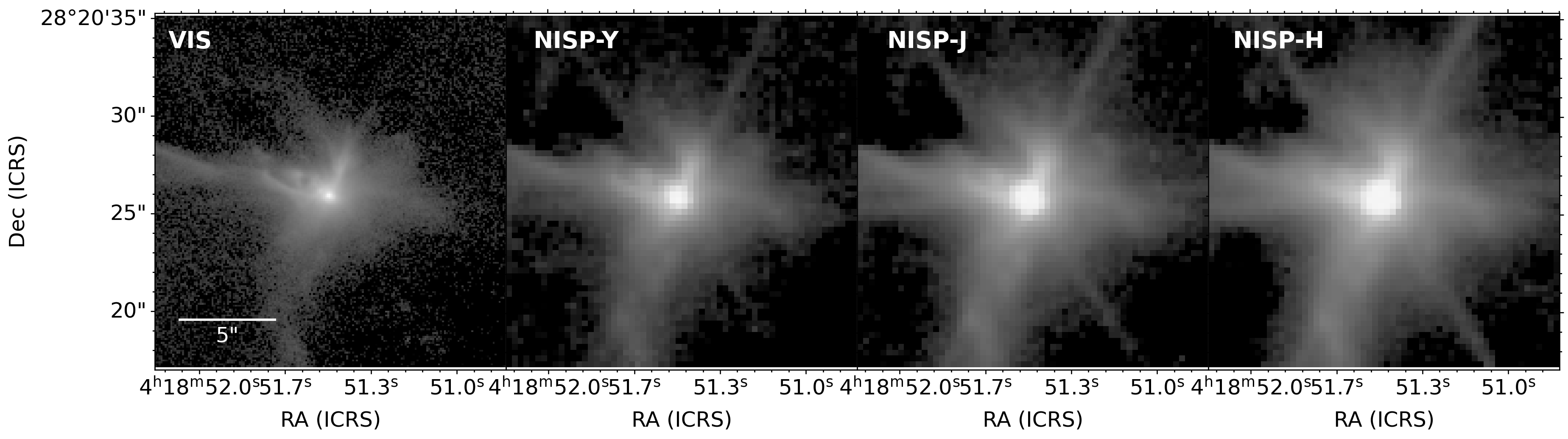}
   \includegraphics[width=0.45\textwidth]{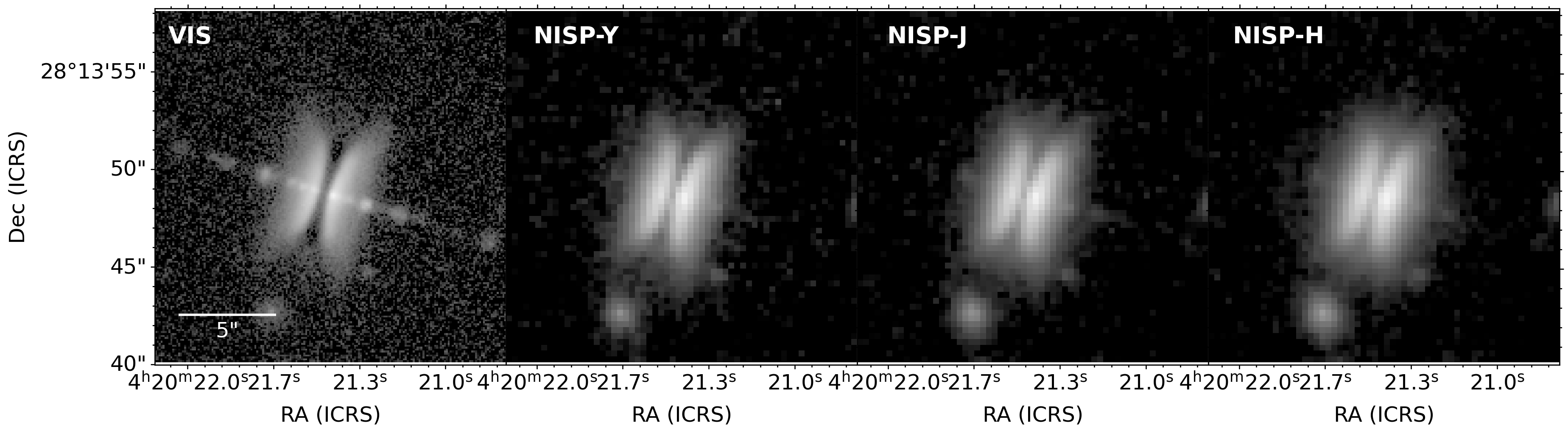}
   \includegraphics[width=0.45\textwidth]{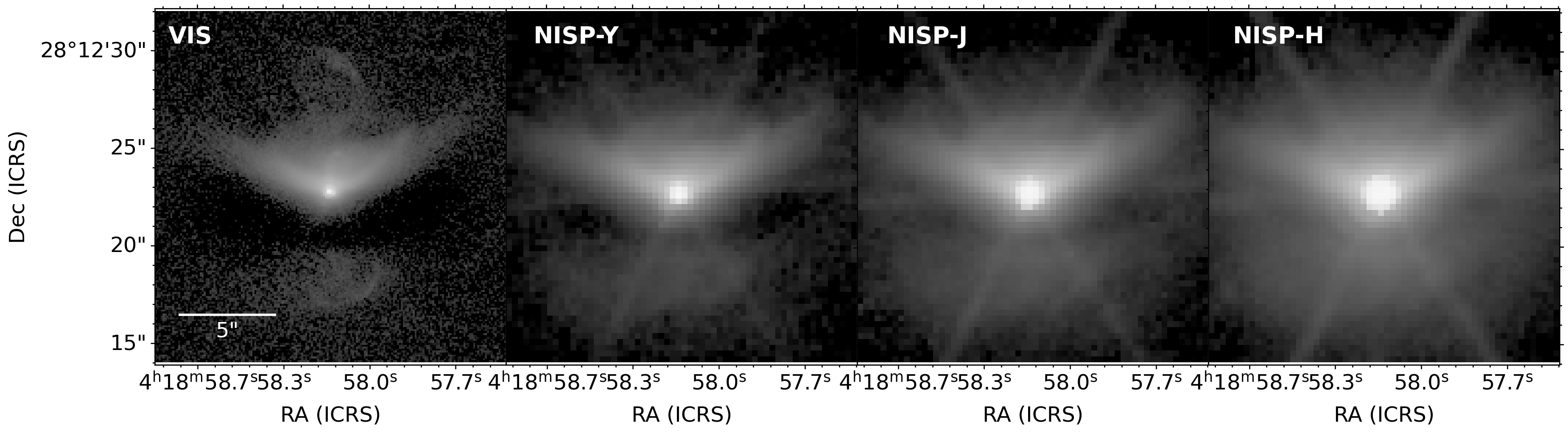}
   \caption{\ec images (log scale) of CoKu Tau 1 (top), 2MASS~J04202144+2813491 (middle), and IRAS04158+2805 (bottom). The source core of CoKu Tau 1 is saturated in the NISP images. The  edge-on disk and the collimated jet around 2MASS~J04202144+2813491 are clearly resolved. A bipolar outflow is observed for the first time around IRAS04158+2805.}
   \label{fig:landmark}%
    \end{figure}

The \ec  images include some famous young stellar objects. We highlight three of them here in Fig.~\ref{fig:landmark}. The exquisite resolution and sensitivity to extended emission of \ec  allowed us to resolve fine structures of their proto-planetary disks, envelopes, outflows and jets. The collimated jets launched by 2MASS~J04202144+2813491 \citep{Luhman2009} are remarkably well detected and resolved in the VIS image, and the knot motion is clearly visible when compared to the  discovery HST images of \citet{Duchene2014}. 

Two lobes are visible in the VIS image of the binary \object{IRAS04158+2805} \citep{Kenyon1990,Ragusa2021}, as illustrated in Fig.~\ref{fig:landmark}. These strongly resemble the bipolar outflow observed in HST images of the pre–main-sequence binary XZ Tauri \citep{Krist2008} and are the first detection of outflows like this around IRAS04158+2805. The outflow seems to be roughly aligned with the jet detected in H$\alpha$ images by \citet{Ragusa2021}. The 0\farcs188 binary is not resolved in the \ec images, possibly because the central star is saturated in the near-infrared images.

\subsection{Proto-brown dwarf candidates}

\begin{figure}
   \centering
    \includegraphics[width=0.45\textwidth]{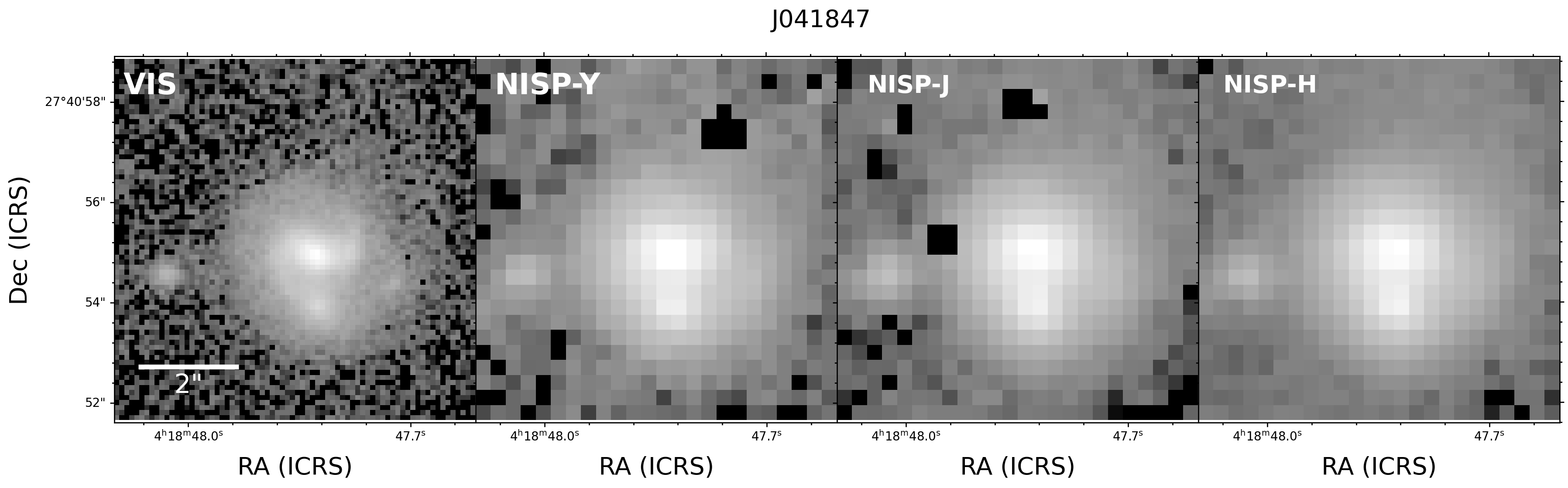}
    \includegraphics[width=0.45\textwidth]{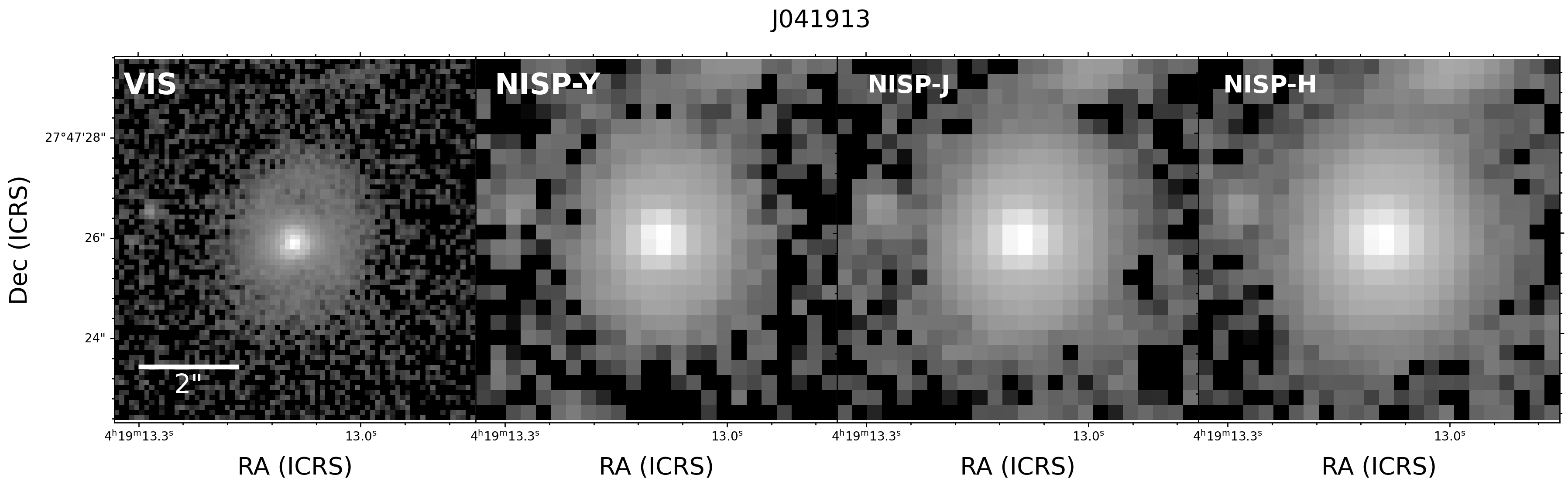}
      \includegraphics[width=0.45\textwidth]{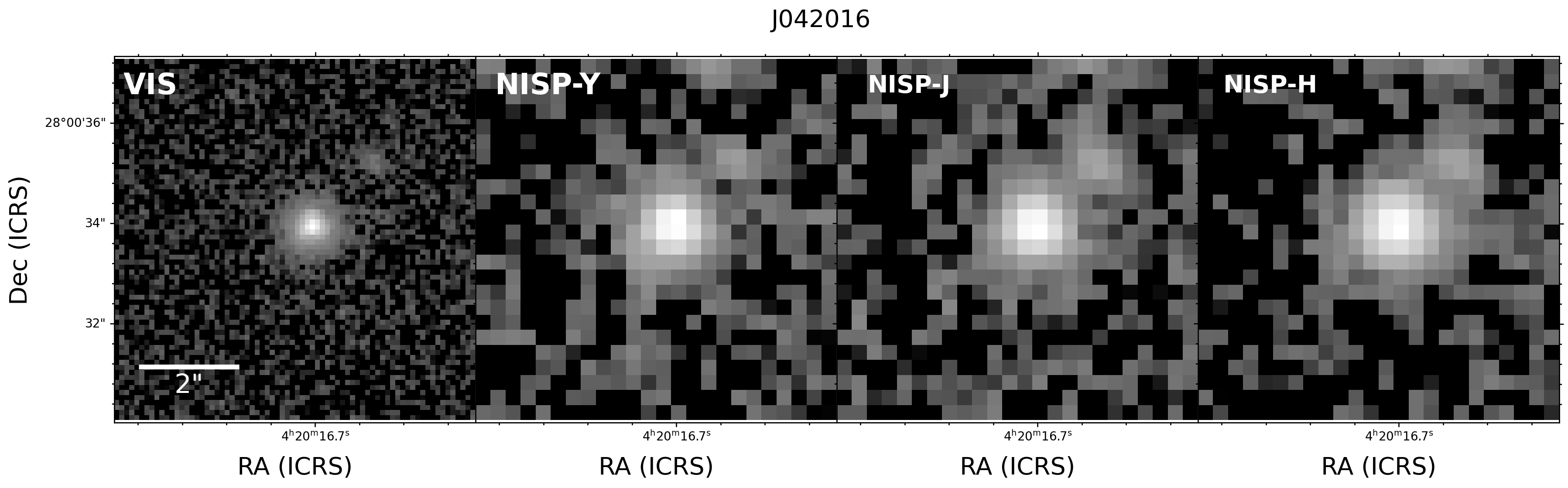}
   \includegraphics[width=0.45\textwidth]{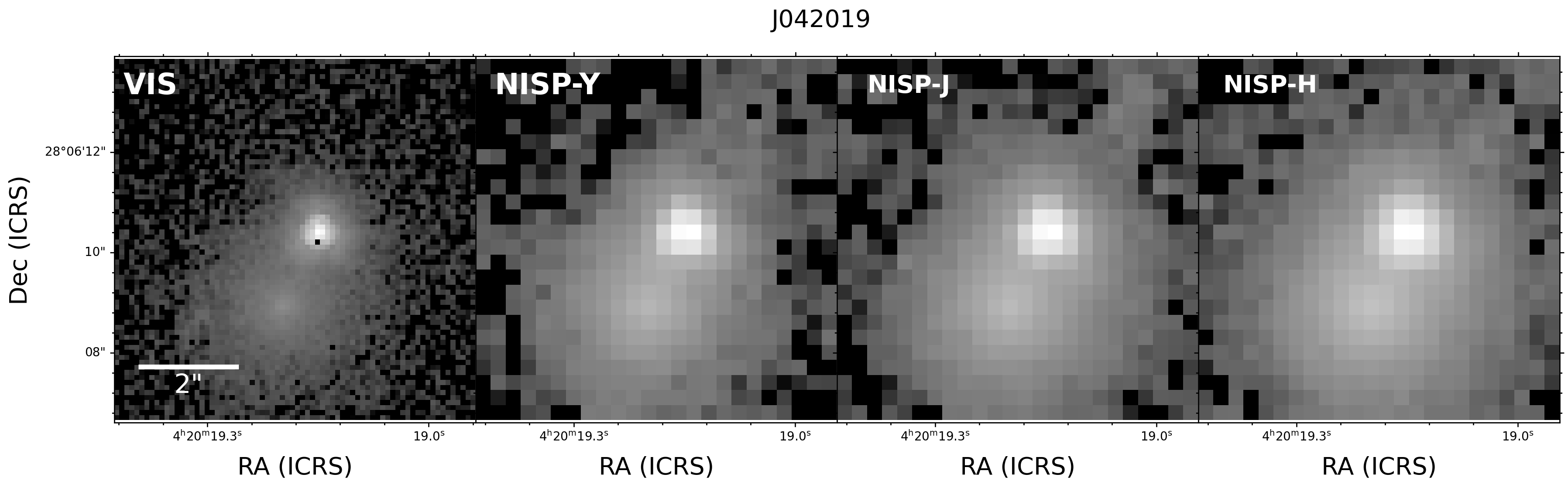}
    \includegraphics[width=0.45\textwidth]{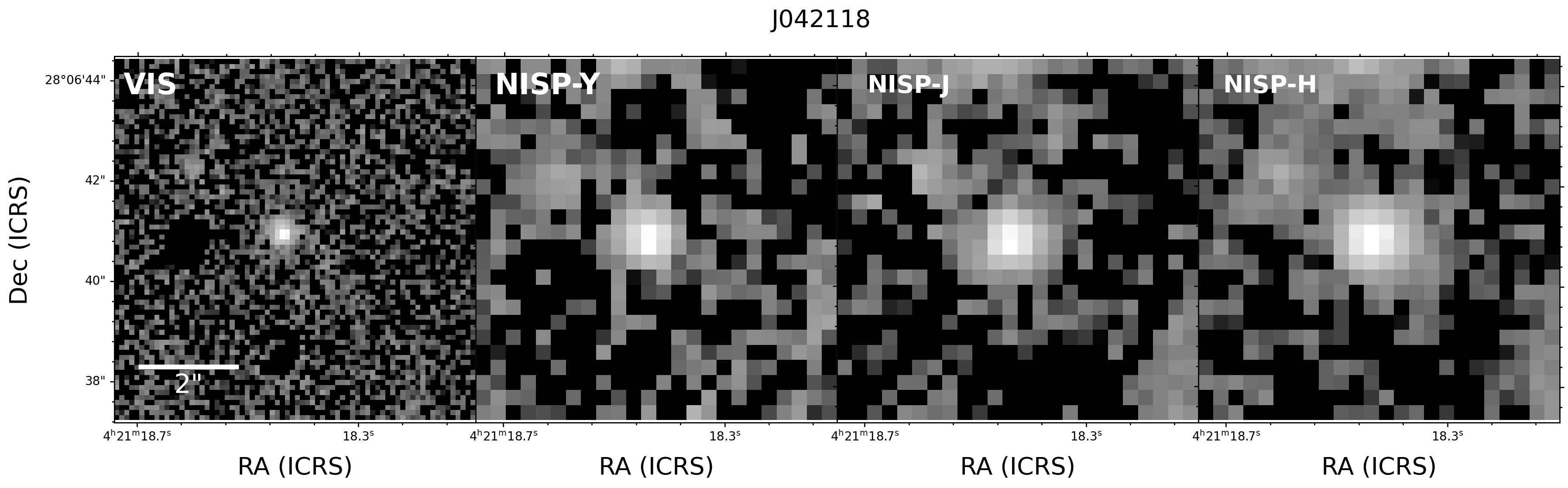}
   \includegraphics[width=0.45\textwidth]{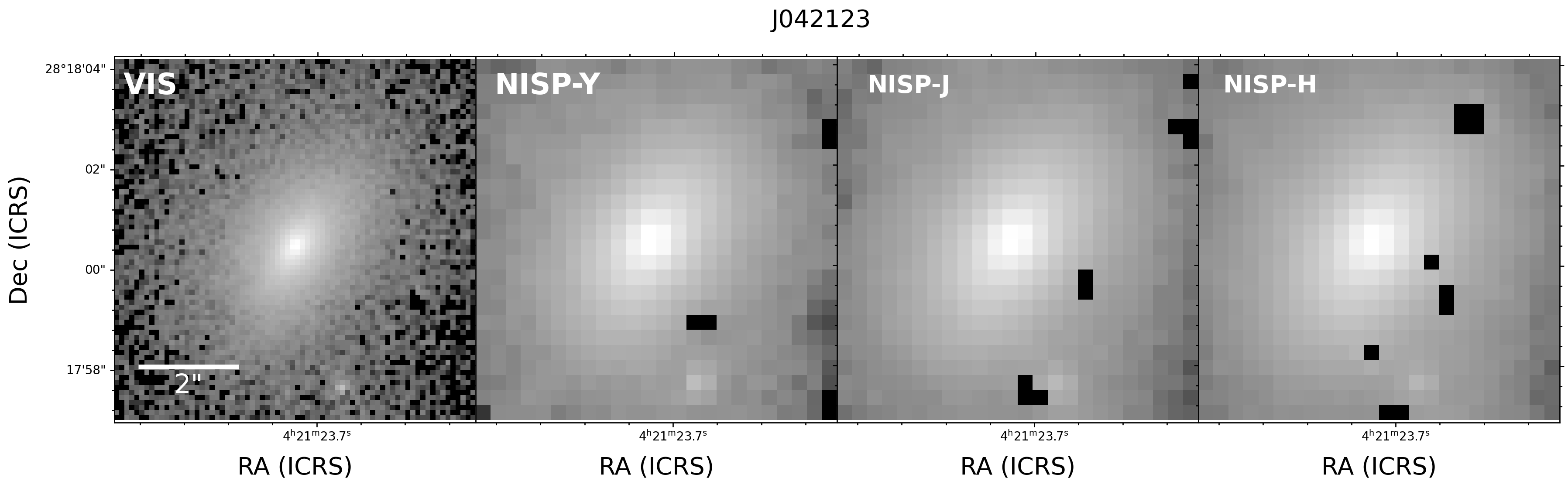}
   \caption{\ec images (log scale) of proto-brown dwarf candidates from \citet{Palau2012} and \citet{Morata2015}. J041847 is clearly resolved as a spiral galaxy, possibly a galaxy merger. J041913 shows some extended emission that suggests that it is a galaxy. J042016, J042019, and J042118 have a neighbor within 3\arcsec\, that might cause the radio emission. J042123 is resolved as an elliptical galaxy.}
   \label{fig:protoBD1}%
    \end{figure}

\begin{figure}
   \centering
   \includegraphics[width=0.45\textwidth]{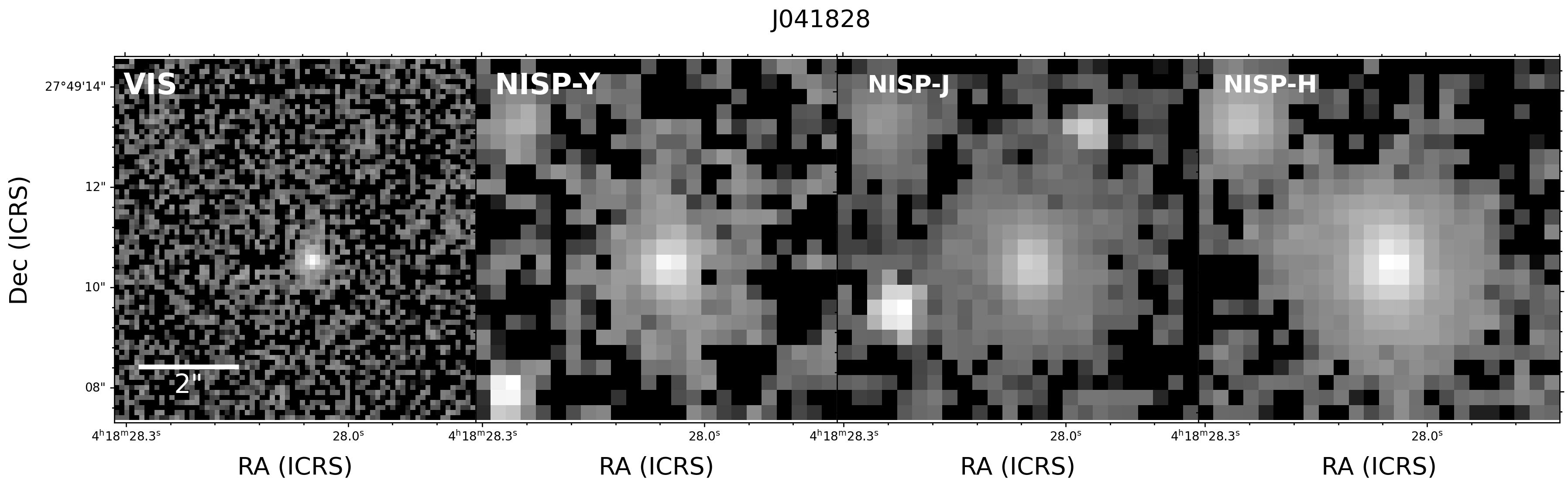}
   \includegraphics[width=0.45\textwidth]{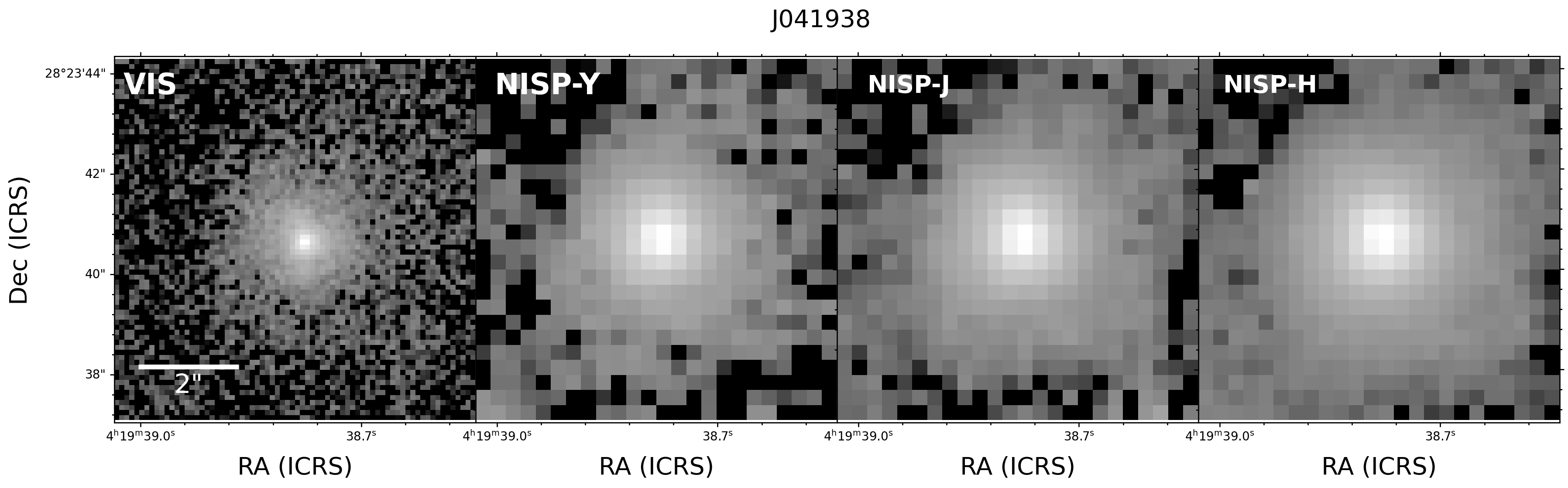}
   \caption{\ec images (log scale) of the proto-brown dwarf candidates J041828 and J041938 from \citet{Palau2012} and \citet{Morata2015}.}
   \label{fig:protoBD2}%
    \end{figure}

Eight of the 12 proto-brown dwarf candidates identified by \citet{Palau2012} and \citet{Morata2015} fall within the \ec field of view. We inspected their VIS and NISP images, as illustrated in Fig.~\ref{fig:protoBD1} and \ref{fig:protoBD2}. The unique sensitivity and spatial resolution of \ec allowed us to test the nature of these candidates. Several objects are clearly resolved as galaxies in the VIS and/or NISP images, as seen in Fig.~\ref{fig:protoBD1}, including J041847 and J042123 \citep[the latter having been already classified as a galaxy based on its radio spectral index by][]{Morata2015}. Source J041913 is also resolved in the VIS image, although its morphology could be compatible with an envelope. Other proto-brown dwarf candidates have a nearby visual companion less than 3\arcsec\, away, which might cause the radio emission we used to identify these candidates given their typical beam size of $\sim$2\arcsec. Two objects (J041828 and J041938) are not resolved in any of the images and are therefore good proto-brown dwarf candidates, as shown in Fig.~\ref{fig:protoBD2}. Table~\ref{tab:protobds} gives a summary of the proto-brown dwarf candidate properties. 

\begin{table}
\centering
\small
\caption{Proto-brown dwarf candidates in the \ec field of view \label{tab:protobds}}

\begin{tabular}{lccl}\hline\hline
Name   & RA (J2000)        & Dec (J2000) & Comment \\
\hline
 J041828 & 04:18:28.08 & 27:49:10.9 & point source  \\ 
 J041847 & 04:18:47.84 & 27:40:55.3 & spiral galaxy / merger   \\ 
 J041913 & 04:19:13.10 & 27:47:26.0 & galaxy?   \\ 
 J041938 & 04:19:38.77 & 28:23:40.7 & point source   \\ 
 J042016 & 04:20:16.70 & 28:00:33.7 & source at $\sim$2\arcsec   \\ 
 J042019 & 04:20:19.20 & 28:06:10.3 & galaxy at $\lesssim$2\arcsec   \\ 
 J042118 & 04:21:18.43 & 28:06:40.8 & source at $\sim$2.5\arcsec    \\ 
 J042123 & 04:21:23.70 & 28:18:00.6 & elliptical galaxy   \\ 
\hline
\end{tabular}
\end{table}

\section{Conclusions}
We presented \ec Early Release Observations of the LDN~1495 molecular cloud in the Taurus star-forming region. We complemented the \ec observations with deep wide-field ground-based images obtained at various facilities over the past 20 years, as well as with newly reprocessed \spitz deep stacks. We selected members based on rather conservative criteria on their morphology, their proper motion, and their positions in nine color-magnitude diagrams and obtained a list of 15 ultracool dwarf candidate members, including 9 new objects and 6 previously confirmed members. The luminosities and colors of the 9 new candidates are consistent with  those of objects with spectral types ranging from late-M to early-T and estimated masses in the range between $\sim$1 and $\sim$15~M$_{\rm Jup}$, according to the  \citet{Chabrier2023} evolutionary models. 

The contamination rate is anticipated to be low given the stringent selection criteria applied here. However, spectroscopic observations are necessary to confirm the nature and cluster membership of these sources, as they could potentially be embedded more massive members.

This study is limited, in particular by the reduced sensitivity and accuracy of our ground-based observations compared to space-based facilities such as \ec, and we relied on theoretical isochrones that currently imperfectly reproduce observations. The current sample is therefore likely both biased and incomplete. With its exquisite astrometric accuracy, achieving a median internal dispersion of $\lesssim$2~mas per epoch \citep{Cuillandre2024}, a second epoch of \ec observations would enable far more accurate proper motion measurements for all the sources in the images. Based on this, we could conclusively determine the membership and nature of these candidate members, and a second observation epoch would facilitate an unbiased and complete census of this region down to sub-Jupiter masses. Nevertheless, this analysis provides a robust and coherent framework for identifying very low-mass members in stellar associations using \ec data, and for the first strong T-dwarf candidates in the Taurus star-forming region.

\begin{acknowledgements} We thank our anonymous referee for their timely and constructive report, which has helped improve this manuscript.
Funding for M{\v Z} and ELM was provided by the European Union (ERC Advanced Grant, SUBSTELLAR, project number 101054354). DB and NH have been funded by the Spanish grants MCIN/AEI/10.13039/501100011033 PID2019-107061GB-C61 and PID2023-150468NB-I00. JO acknowledge financial support from "Ayudas para contratos postdoctorales de investigación UNED 2021".
This work has made use of the Early Release Observation (ERO) data from the European Space Agency (ESA) mission Euclid, available at \url{https://euclid.esac.esa.int/dr/ero/}.
We are grateful to A. Schneider, K. Luhman, M. Liu, M. Bonnefoy for providing us with the IR spectra of young ultracool dwarfs presented in the various color-magnitude diagrams. We thank I. Baraffe and M. Phillips for providing the ATMO 2020 and Chabrier+2023 models in the \ec filter set.
This work has made use of data from the European Space Agency (ESA) mission {\it Gaia} (\url{https://www.cosmos.esa.int/gaia}), processed by the {\it Gaia} Data Processing and Analysis Consortium (DPAC, \url{https://www.cosmos.esa.int/web/gaia/dpac/consortium}). Funding for the DPAC has been provided by national institutions, in particular the institutions participating in the {\it Gaia} Multilateral Agreement.
This research has made use of the NASA/IPAC Infrared Science Archive, which is funded by the National Aeronautics and Space Administration and operated by the California Institute of Technology.
This research has made use of the Spanish Virtual Observatory (\url{https://svo.cab.inta-csic.es}) project funded by MCIN/AEI/10.13039/501100011033/ through grant PID2020-112949GB-I00.
This work has benefited from The UltracoolSheet at \url{http://bit.ly/UltracoolSheet}, maintained by Will Best, Trent Dupuy, Michael Liu, Rob Siverd, and Zhoujian Zhang.
Based in part on data collected at Subaru Telescope which is operated by the National Astronomical Observatory of Japan and obtained from the SMOKA, which is operated by the Astronomy Data Center, National Astronomical Observatory of Japan. 
This research has made use of the VizieR catalogue access tool, CDS, Strasbourg, France. The original description of the VizieR service was published in \citet{vizier}.
This work made use of GNU Parallel \citep{Tange2011a}, astropy \citep{astropy2013,astropy2018}, Topcat \citep{Topcat}, matplotlib \citep{matplotlib}, Bokeh \citep{Bokeh}, Plotly \citep{plotly}, Numpy \citep{numpy}, APLpy \citep{aplpy2012,aplpy2019}.
\end{acknowledgements}

%
%

\bibliographystyle{aa} 

\begin{appendix}

\section{Complementary ground-based observations \label{sec-groundobs}}
Table~\ref{tab:observations} gives an overview of the ground-based wide-field instruments used in this study to complement the \ec  data.

\begin{table*}
  \small
\caption{Ground-based instruments used in this study \label{tab:observations}}

\begin{tabular}{lccccccclcc}\hline\hline
Observatory   & Instrument        & Filters              & Platescale   & Field of view & Epoch Min./Max.  & Ref. \\
                      &                          &                         & [\arcsec pixel$^{-1}$] &               &       &                  &                     &                           &      \\
\hline
CFHT & UH12K & \textit{I} & 0.20 & 42\arcmin$\times$28\arcmin & 2000--2003 &  (1) \\
CFHT   & MegaCam   & \textit{r,i,z}  & 0.18 & 1\degr$\times$1\degr & 2003-2019  &  (2) \\
UKIRT &  WFCAM & \textit{Z,Y,J,H,Ks} & 0.4 & 40\arcmin$\times$40\arcmin  & 2005--2016 &  (3)\\
Subaru & HSC & \textit{r,i,z,y} & 0.17 & 45\arcmin\, radius & 2015--2016 &  (4) \\
KPNO 4m & NEWFIRM & \textit{J,H,Ks} & 0.40 & 14\arcmin$\times$14\arcmin & 2013, 2016 &  (5) \\
CAHA 3.5 & $\Omega$2000 & \textit{H,Ks} & 0.45 &  15\arcmin$\times$15\arcmin & 2011--2016 &  (6) \\
JAST & T80cam & \textit{i,z} & 0.55 & 85\arcmin$\times$85\arcmin  & 2016, 2019 &  (7) \\
Subaru & SuprimeCam & \textit{i,z} & 0.2 & 34\arcmin$\times$27\arcmin & 2004--2009 &  (8) \\
CFHT & WIRCam & \textit{J,H,W} & 0.3 & 10.5\arcmin$\times$10.5\arcmin & 2015--2018 &  (9) \\
\hline
\end{tabular}
\tablebib{(1) \citet{Cuillandre2000}, (2) \citet{Boulade2003}, (3) \citet{UKIRT_WFCAM}, (4) \citet{HSC}, (5) \citet{NEWFIRM}, (6) \citet{O2000}, (7) \citet{OAJ}, (8) \citet{Suprimecam}, (9) \citet{WIRCam}}

\end{table*}

\section{Benchmark ultracool dwarf spectro-photometry in the Euclid and \textit{Spitzer} filters \label{sec:spectrophoto}}

\citet{NISP_FILTERS} computed color transformations from or to popular systems (2MASS, UKIDSS,...) for a variety of spectral energy distributions including field very low mass stars and brown dwarfs. 
Rather than relying on these transformations we decided to compute the photometry of known ultracool dwarfs in the  NISP filters using calibrated near-infrared spectra from the literature and including: 
\begin{itemize}
\item 3 young late-L dwarfs members of the IC348 cluster observed with the JWST and its NIRSpec spectrograph \citep{Luhman2024}
\item A set of 23 young T dwarfs members of nearby moving groups observed with the NASA Infrared Telescope Facility (IRTF) with the SpeX spectrograph  \citep{Zhang2021}
\item The young planet VHS 1256-1257~b spectrum obtained with the JWST and its NIRSpec spectrograph \citep{Miles2023}
\item The young planetary mass companion 2MASS1207~b spectrum obtained with the JWST and its NIRSpec spectrograph \citep{Luhman2023a}
\item the young T3.5 dwarf companion Gu Psc b spectrum presented in \citet{Naud2014}
\item The young L7 dwarf WISEA~J114724.10-204021.3 spectrum presented in \citet{Schneider2016}
\item the young L7 dwarf PSO J318.5338-22.8603 spectrum reported in \citet{Liu2013}
\item the list of all ultracool dwarfs classified as low or intermediate gravity in the IRTF/SpeX library \citep{SPLAT}
\item the SpeX  spectral standard library of \citet{SPLAT} for field (old) ultracool dwarfs
\end{itemize}

We also include the \ec  photometry presented in \citet{Martin2024} of confirmed young members from the $\sigma$-Orionis cluster from \citet{Pena2012}.

In all cases the SpeX Prism Library Analysis Toolkit \citep[SPLAT;][]{SPLAT} was used to derive the spectro-photometry from the calibrated spectra. 

\subsection{Spitzer IRAC 1 and 2 spectrophotometry}

The photometry in the \spitz IRAC 1 and 2 filters was already available for the 23 young T dwarfs from \citet{Zhang2021}, for some of the $\sigma-$Orionis members, for Gu Psc b, WISEA~J114724.10-204021.3, PSO J318.5338-22.8603, as well as a significant fraction of the low or intermediate gravity ultracool dwarfs from the IRTF/SpeX library. We derived the IRAC1 and 2 spectro-photometry of the 3 young late-L dwarfs members of IC348 as well as VHS 1256-1257~b from the JWST NIRSpec spectra  using SPLAT and the Spitzer transmission curves as described above. 

\begin{figure}
   \centering
   \includegraphics[width=0.4\textwidth]{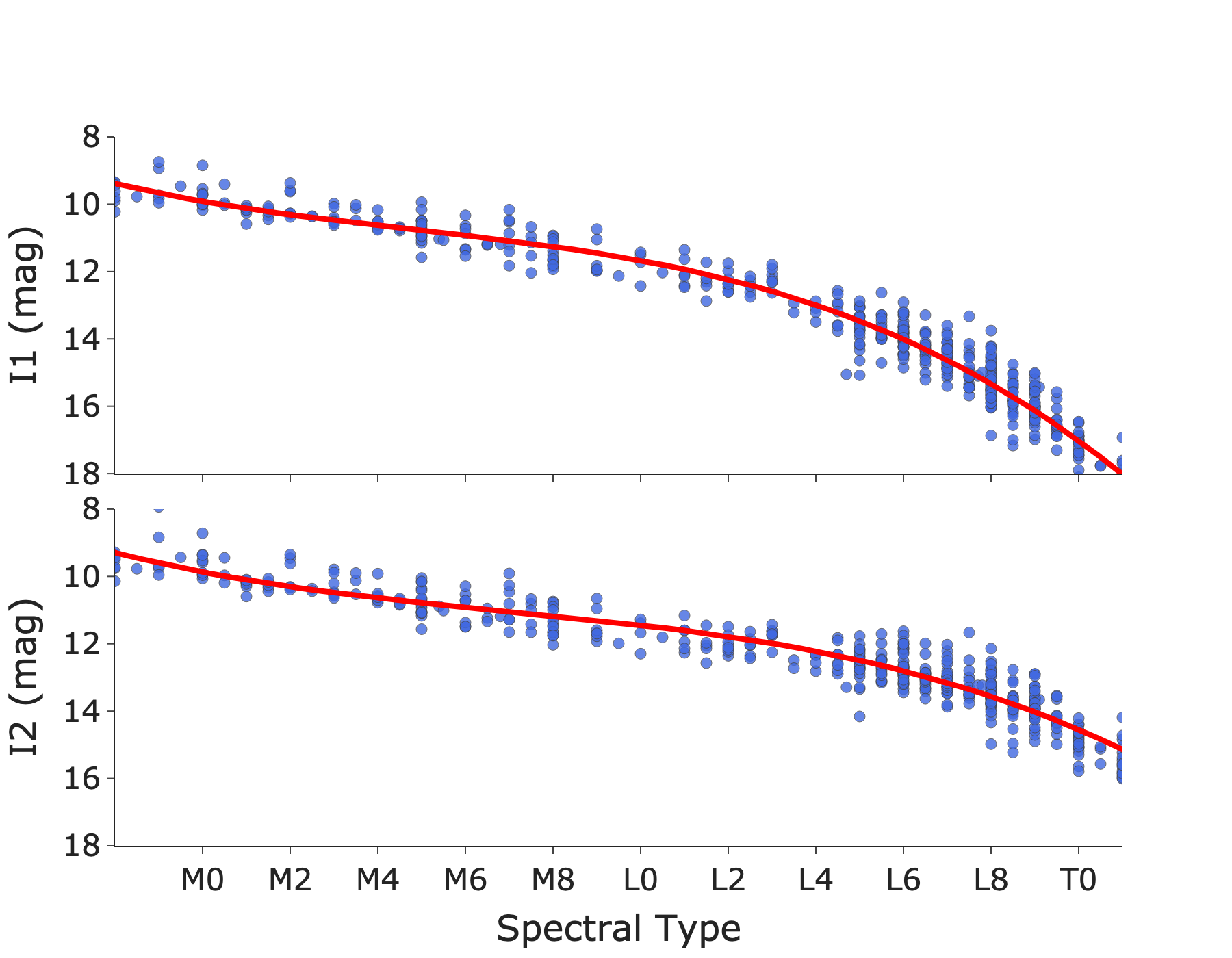}
   \cprotect\caption{Absolute IRAC1 and IRAC2 magnitudes vs spectral type for ultracool dwarfs from  the UltracoolSheet v2.0 \citep{UltracoolSheet}. The  third order Chebyshev polynomial fit is represented in red.}
   \label{fig:spt_i1_i2}%
    \end{figure}

We also augmented the SpeX ultracool spectral standard library of \citet{SPLAT} in the IRAC 1 and 2 filters. For that purpose we fitted a third order Chebyshev polynomial to the IRAC1 and IRAC2 absolute magnitudes as a function of spectral type for all the M, L and T dwarfs with parallax measurements reported in the UltracoolSheet v2.0 \citep{UltracoolSheet}. The result is presented in Fig.~\ref{fig:spt_i1_i2} and we added the corresponding values to the SpeX ultracool spectral standards spectrophotometry.

\begin{figure}
   \centering
   \includegraphics[width=0.49\textwidth]{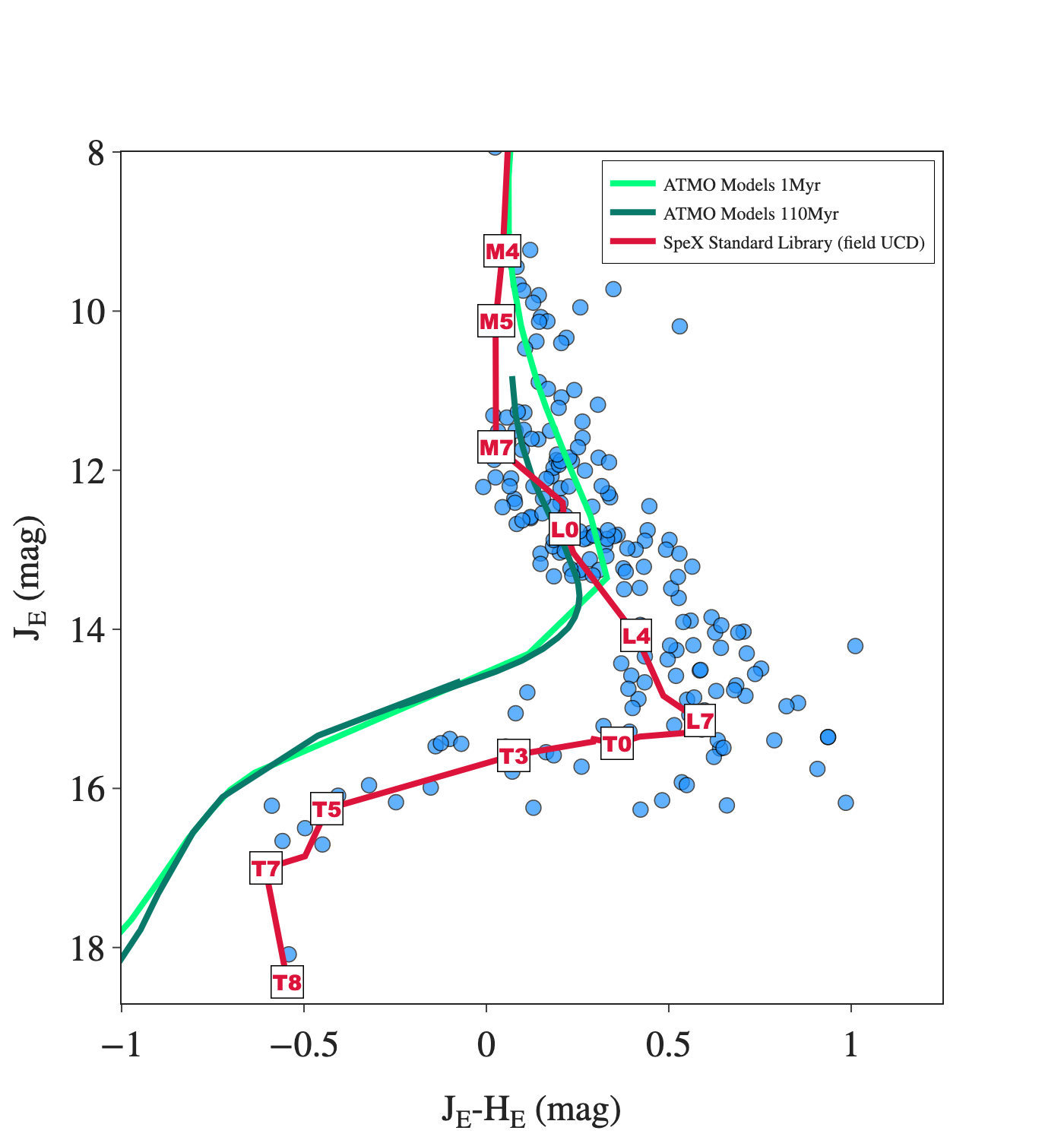}
   \cprotect\caption{Absolute (\je, \je-\he) color-magnitude diagram for the known young or low-gravity ultracool dwarfs from Table~\ref{tab:known-UCD} (blue dots). The \citet{Chabrier2023} models for 1 and 110Myr are represented as light and dark green lines, respectively. The SPLAT standard spectral sequence for older field ultracool dwarfs presented in Table~\ref{table:spex_std} is represented in red, and the corresponding spectral types are indicated.}
   \label{fig:cmd_known_ucd}%
    \end{figure}

Table~\ref{table:spex_std} gives the sequence in the five filters from M0 to T8. The \ec and Spitzer spectrophotometry for all the young and low-gravity ultracool dwarfs computed that way is reported in Table~\ref{tab:known-UCD} and provides a useful reference for all future studies of young ultracool dwarfs using \ec data. Figure~\ref{fig:cmd_known_ucd} shows the sequence and all the known ultracool dwarfs in a (\je,\je-\he) color-magnitude diagram

\begin{table*}
\caption{Absolute magnitudes of known young or low-gravity ultracool objects from the literature in the \ec and z filters. \label{tab:known-UCD}}
  \scriptsize
\begin{tabular}{lccccccccc}
\hline\hline
  Object &  RA (J2000) &  Dec (J2000) &  SpT &  \textit{z} &  \ye & \je & \he &  IRAC1 &  IRAC2 \\
         & deg & deg &          & mag & mag & mag & mag & mag & mag \\
\hline
  2MASP J0345432+254023 & 56.42983 & 25.67314 & L1 & 14.78 & 13.14 & 12.71 & 12.49 &  &  \\
2MASS J00193927-3724392 & 4.91362 & -37.41089 & L3 &  & 13.82 & 13.21 & 12.78 &  &  \\
2MASS J00332386-1521309 & 8.34942 & -15.35858 & L1 & 16.59 & 15.01 & 14.26 & 13.74 & 10.73 & 10.67 \\
\multicolumn{10}{c}{...} \\
\hline
\end{tabular}
\tablefoot{The entire table is available in electronic form}
\end{table*}

\begin{table}
\caption{Ultracool spectral standards sequence in the \ec  NISP and Spitzer IRAC1 and 2 filters}             
\label{table:spex_std}
\centering
\small
\begin{tabular}{c c c c c c}
\hline\hline           
Specral Type & \ye & \je & \he & \textit{I1} & \textit{I2} \\
\hline   
M0 & 7.03 & 6.86 & 6.81 & 4.82 & 4.77 \\
M1 & 7.50 & 7.38 & 7.36 & 5.68 & 5.58 \\
M2 & 7.90 & 7.72 & 7.66 & 6.44 & 6.32 \\
M4 & 9.44 & 9.24 & 9.19 & 7.70 & 7.56 \\
M5 & 10.32 & 10.13 & 10.10 & 8.22 & 8.08 \\
M7 & 11.89 & 11.7 & 11.67 & 9.05 & 8.94 \\
M9 & 12.81 & 12.41 & 12.2 & 9.67 & 9.60 \\
L0 & 13.17 & 12.74 & 12.52 & 9.92 & 9.87 \\
L1 & 13.48 & 13.04 & 12.80 & 10.13 & 10.10 \\
L4 & 14.64 & 14.08 & 13.67 & 10.63 & 10.64 \\
L6 & 15.47 & 14.84 & 14.35 & 10.93 & 10.92 \\
L7 & 15.83 & 15.15 & 14.57 & 11.08 & 11.04 \\
L8 & 15.93 & 15.29 & 14.71 & 11.26 & 11.17 \\
L9 & 15.91 & 15.35 & 14.93 & 11.45 & 11.30 \\
T0 & 15.91 & 15.44 & 15.08 & 11.67 & 11.45 \\
T1 & 15.77 & 15.37 & 15.09 & 11.93 & 11.61 \\
T2 & 15.83 & 15.41 & 15.11 & 12.23 & 11.78 \\
T3 & 15.93 & 15.58 & 15.50 & 12.59 & 11.99 \\
T5 & 16.61 & 16.26 & 16.70 & 13.47 & 12.49 \\
T6 & 17.30 & 16.85 & 17.35 & 14.01 & 12.80 \\
T7 & 17.47 & 17.01 & 17.61 & 14.63 & 13.16 \\
T8 & 19.04 & 18.43 & 18.98 & 15.33 & 13.56 \\
\hline 
\end{tabular}
\end{table}

\section{\textit{Euclid} filter properties \label{sec-euclidfilters}}

Table~\ref{table:euclidfilters} shows the \ec filters properties used in this study, obtained or computed using the values listed in the Spanish Virtual Observatory Filter Profile Service \citep[][\url{https://svo.cab.inta-csic.es}]{SVOFilters}.

\begin{table}
\caption{Properties of the \textit{Euclid} filters used in this article.}             
\label{table:euclidfilters}
\centering                 
\begin{tabular}{c c c c c }
\hline\hline           
Filter & $\lambda_{\rm cen}$ & width & m$_{\rm AB}$- m$_{\rm Vega}$ & A$_{\lambda}$/$A_{\rm V}$ \\
       & (nm) & (nm) & (mag)      & \\    
\hline   
\ie &  715 & 355 & 0.268 & 0.717 \\
\ye &  1080.9 &  262.7 & 0.694 &  0.382 \\
\je &  1367.3 &  399.4 & 1.058 & 0.263 \\
\he &  1771.4 &  499.9 & 1.492 & 0.171 \\
\hline                                   
\end{tabular}
\end{table}

\end{appendix}

\end{document}